\documentclass[twocolumn]{aastex701}

\newcommand\xmm{\textit{XMM-Newton}}
\newcommand\chandra{\textit{Chandra}}

\newcommand\nustar{\textit{NuSTAR}}

\newcommand\xrism{\textit{XRISM}}

\newcommand\delcstat{$\Delta$C-stat}
\newcommand\pcm{cm$^{-2}$}

\newcommand\logxi{$\log (\xi$/erg~cm~s$^{-1})$}

\newcommand\nh{N$_{\rm H}$}

\usepackage{multirow}

\begin{document}

\title{\xrism/Resolve observations of Hercules X-1: vertical structure and kinematics of the disk wind}

\author[orcid=0000-0003-4511-8427,sname='Kosec']{Peter Kosec}
\affiliation{Center for Astrophysics | Harvard \& Smithsonian, Cambridge, MA, USA}
\email[show]{peter.kosec@cfa.harvard.edu}

\author[orcid=0000-0003-2663-1954,sname='']{Laura Brenneman}
\affiliation{Center for Astrophysics | Harvard \& Smithsonian, Cambridge, MA, USA}
\email[]{peter.kosec@cfa.harvard.edu}

\author[orcid=0000-0003-0172-0854,sname='']{Erin Kara}
\affiliation{MIT Kavli Institute for Astrophysics and Space Research, Massachusetts Institute of Technology, Cambridge, MA 02139, USA}
\email[]{peter.kosec@cfa.harvard.edu}

\author[orcid=0000-0003-1244-3100,sname='']{Teruaki Enoto}
\affiliation{Department of Physics, Graduate School of Science, Kyoto University, Kitashirakawa Oiwake-cho, Sakyo-ku, Kyoto, 606-8502, Japan}
\email[]{peter.kosec@cfa.harvard.edu}  

\author[orcid=0009-0006-7889-6144,sname='']{Takuto Narita}
\affiliation{Department of Physics, Graduate School of Science, Kyoto University, Kitashirakawa Oiwake-cho, Sakyo-ku, Kyoto, 606-8502, Japan}
\email[]{peter.kosec@cfa.harvard.edu}

\author[orcid=0009-0007-8032-3641,sname='']{Koh Sakamoto}
\affiliation{Department of Physics, Graduate School of Science, Kyoto University, Kitashirakawa Oiwake-cho, Sakyo-ku, Kyoto, 606-8502, Japan}
\email[]{peter.kosec@cfa.harvard.edu}  

\author[orcid=0000-0003-1498-1543,sname='']{Rüdiger Staubert}
\affiliation{Institut für Astronomie und Astrophysik, Universität Tübingen, Sand 1, D-72076 Tübingen, Germany}
\email[]{peter.kosec@cfa.harvard.edu}

\author[orcid=0000-0001-5852-6740,sname='']{Francesco Barra}
\affiliation{Università degli Studi di Palermo, Dipartimento di Fisica e Chimica, via Archirafi 36, I-90123 Palermo, Italy}
\email[]{peter.kosec@cfa.harvard.edu}  

\author[orcid=0000-0002-9378-4072,sname='']{Andrew Fabian}
\affiliation{Institute of Astronomy, Madingley Road, Cambridge, CB3 0HA, UK}
\email[]{peter.kosec@cfa.harvard.edu}

\author[orcid=0000-0003-2869-7682,sname='']{Jon M. Miller}
\affiliation{Department of Astronomy, University of Michigan, Ann Arbor, MI 48109, USA}
\email[]{peter.kosec@cfa.harvard.edu}  

\author[orcid=0000-0003-2532-7379,sname='']{Ciro Pinto}
\affiliation{INAF—IASF Palermo, Via U. La Malfa 153, I-90146 Palermo, Italy}
\email[]{peter.kosec@cfa.harvard.edu}

\author[orcid=0000-0002-5359-9497,sname='']{Daniele Rogantini}
\affiliation{Department of Astronomy and Astrophysics, The University of Chicago, Chicago, IL 60637, USA}
\email[]{peter.kosec@cfa.harvard.edu}

\author[orcid=0000-0001-5819-3552,sname='']{Dominic Walton}
\affiliation{Centre for Astrophysics Research, University of Hertfordshire, UK}
\email[]{peter.kosec@cfa.harvard.edu}  

\author[orcid=0009-0003-9261-2740,sname='']{Yutaro Nagai}
\affiliation{Department of Physics, Graduate School of Science, Kyoto University, Kitashirakawa Oiwake-cho, Sakyo-ku, Kyoto, 606-8502, Japan}
\email[]{peter.kosec@cfa.harvard.edu}




\begin{abstract}


X-ray binary accretion disk winds can carry away a significant fraction of the matter transferred from the companion and hence strongly affect the accretion flow and the long-term evolution of the binary. However, accurate mass outflow rate measurements are challenging due to uncertainties in our understanding of the 3D wind structure. Most studies employ absorption line spectroscopy that only gives us a single sightline through the wind streamlines. Hercules X-1 is a peculiar X-ray binary which allows us to avoid this issue, as its warped, precessing accretion disk naturally presents a range of sightlines through the vertical structure of its disk wind. Here we present the first results from a large, coordinated campaign on Her X-1 led by the new \xrism\ observatory (with an exposure of 210 ks) and supported by \xmm, \nustar\ and \chandra. We perform a time-resolved analysis and constrain the wind properties. With \xrism/Resolve, we directly detect the Her X-1 orbital motion with an amplitude of 170 km/s in the evolution of the wind velocity. After correcting for this effect, we observe an increase in wind velocity from 250 km/s to 600 km/s as the wind rises to greater heights above the disk. The wind column density decreases with increasing height, as expected, but its ionization parameter \logxi\ evolves only weakly from 3.65 to 3.9 as the wind expands away. Additionally, we detect a new orbital dependence of the wind properties, revealing a likely second component that appears only briefly after the eclipse by the secondary star.

\end{abstract}

\keywords{\uat{Accretion}{14} --- \uat{High Energy astrophysics}{739} --- \uat{Neutron stars}{1108}}


\section{Introduction}

Blueshifted absorption lines imprinted by accretion disk winds are some of the most distinct spectral features of X-ray binaries. First discovered using the ASCA observatory \citep{Ueda+98, Kotani+00}, our understanding of disk winds and their observational signatures has significantly improved over the last $\sim2$ decades thanks to \chandra\ and \xmm\ \citep[e.g.][]{Ueda+04, Trigo+12}. More recently, signatures of disk winds in X-ray binaries were also discovered in the UV, optical and infrared energy bands \citep{Munoz-Darias+19, Sanchez-Sierras+20, Castro-Segura+22}. 

Typically, disk winds in X-ray binaries reach velocities up to 1000 km/s \citep[for a review, see][]{Neilsen+23}, much lower than certain types of wide-angle outflows observed in accreting supermassive black holes \citep[e.g.][]{Tombesi+10} and ultraluminous X-ray sources \citep[e.g.][]{Pinto+16, Kosec+18b} that can exhibit velocities in excess of $10\%$ of the speed of light. Nevertheless, the X-ray binary disk wind mass outflow rates can be of the same order as the mass accretion rates onto the compact object \citep{Lee+02, Kosec+20}. Therefore, by transporting this significant mass as well as angular momentum outside the binary system, they can affect the long-term evolution of X-ray binaries \citep{Tetarenko+18, Gallegos+24}.

Despite being known for over 2 decades, much is still unknown about these phenomena in X-ray binaries, including their exact physical launching mechanism. It is also challenging to accurately quantify their mass outflow rates, and thus constrain the actual impact on their binary systems and neighborhoods. One of the reasons why this is difficult is because disk winds are mainly studied through their X-ray absorption lines, and thus are only sampled along a single line of sight towards the compact X-ray source. X-ray absorption line spectroscopy is the most straightforward way to understand the wind properties such as its column density, ionization parameter and velocity, but it gives us a limited view of the complex 3D structure of the outflow. Statistical studies of samples of X-ray binaries \citep{Ponti+12, Parra+24} suggest that the vertical wind structure is not spherical but equatorial (with a $\sim10^{\circ}$ half-opening angle), but this parameter cannot typically be constrained in any individual source.

Hercules X-1 (hereafter Her X-1) is an X-ray pulsar, an X-ray binary powered by a highly magnetized neutron star rotating every 1.2 seconds \citep{Tananbaum+72}. It is famous for its 35-day cycle of high and low flux states, introduced by a warped, precessing accretion disk that is seen almost edge-on \citep{Katz+73, Gerend+76}. This warped disk precession changes our line-of-sight through the accretion flow, and at certain times of the 35-day precession cycle, outer parts of the disk obscure our view of the inner accretion flow (which produces most of the X-rays). At other times, the inner accretion flow is uncovered, resulting in two high-flux periods during each 35-day cycle named the Main High and the Short High states. 

The disk precession offers a unique and powerful tool to study the 3D properties of the disk wind of Her X-1, as our time-variable sightline intersects the wind streamlines at different locations over the 35-day precession cycle, specifically at different heights above the accretion disk. A schematic showing this situation is shown in Fig. 1 of \citet{Kosec+23a}. According to the model of the Her X-1 warped disk precession by \citet{Scott+00}, the inner and outer rings of the accretion disk precess by 11$^{\circ}$ and 20$^{\circ}$, respectively, over the 35-day precession cycle (this motion is shown from the point of view of the neutron star in Fig. 2 of their paper). A significant part of the wind structure then crosses our line of sight during the 35-day cycle, assuming that the structure is equatorial with a relatively small opening angle \citep[as observed in other X-ray binaries,][]{Ponti+12}.

Thus, we can explore the wind vertical structure through its X-ray absorption lines by sampling the wind properties over a range of precession phases. We previously studied Her X-1 with \xmm\ and \chandra\ and showed that the disk wind is present in X-ray absorption during both the Main and Short High states \citep{Kosec+20, Kosec+23c} and has a velocity of $250-1000$ km/s and an ionization parameter of \logxi\ of $3-4$, consistent with typical X-ray binary disk winds. \citet{Kosec+20} and \citet{Kosec+23a} performed a detailed study of the wind during the Main High state and leveraged the disk precession to produce a 2D map of this outflow, and estimated the mass outflow rate to be as high as $70\%$ of the mass transfer rate from the secondary star.

Typical accretion disk winds detected in X-ray binaries are highly ionized, with ionization parameters around \logxi\ of 4. Some of their strongest observational signatures are therefore in the Fe K energy band, with particularly strong transitions of Fe XXV (at 6.7 keV) and Fe XXVI (at 6.96 keV). The recently launched \xrism\ observatory \citep{Tashiro+22} offers a generational leap in our spectral capability in this energy band. The microcalorimeter Resolve instrument onboard \xrism\ achieves a spectral resolution of 4.5 eV at 6 keV \citep{Ishisaki+22}, which is roughly $\sim10\times$ the spectral resolution of \chandra\ HETG gratings (first order) and $\sim30\times$ the spectral resolution of typical X-ray CCDs such as EPIC onboard \xmm\ at these energies. Therefore, \xrism\ is an outstanding instrument to study the physics of accretion disk winds in X-ray binaries, providing a precision and detail much greater than has previously been possible.

In September 2024, we carried out a large observational campaign on Her X-1 led by \xrism. The aim of the campaign is to study and understand the 3D structure of the Her X-1 disk wind with unprecedented precision \citep{Kosec+24}. To achieve this goal, we observed a significant fraction of a single Main High state of Her X-1 for a duration of $\sim400$ ks (resulting in a $\sim200$ ks clean exposure time after accounting for \xrism's low-earth orbit). In addition to this \xrism\ observation, we also obtained simultaneous observations with \chandra\ (50 ks), \xmm\ (80 ks) and \nustar\ (40 ks), each covering part of the long \xrism\ exposure.

This is the first manuscript describing the 2024 \xrism\ observations of Her X-1. Here we focus on the evolution of the wind properties with Her X-1 precession phase. Thanks to the excellent spectral capabilities of \xrism, we are for the first time able to track the kinematics of the wind and their variations with precession phase. The manuscript is structured as follows. In section \ref{sec:obsdataprep}, we describe the September 2024 observational campaign on Her X-1, and the preparation and reduction procedures for all datasets. In section \ref{sec:results}, we present the results of this study using phenomenological as well as physically motivated spectral fitting approaches. In section \ref{sec:discussion}, we discuss the results and in section \ref{sec:conclusions} we summarize the conclusions of this paper. Throughout the paper, we assume a distance of 6.1 kpc for Her X-1 \citep{Leahy+14}.

\section{Observations and Data Preparation}
\label{sec:obsdataprep}

\subsection{2024 observations of Hercules X-1}

\begin{figure*}
\includegraphics[width=\textwidth]{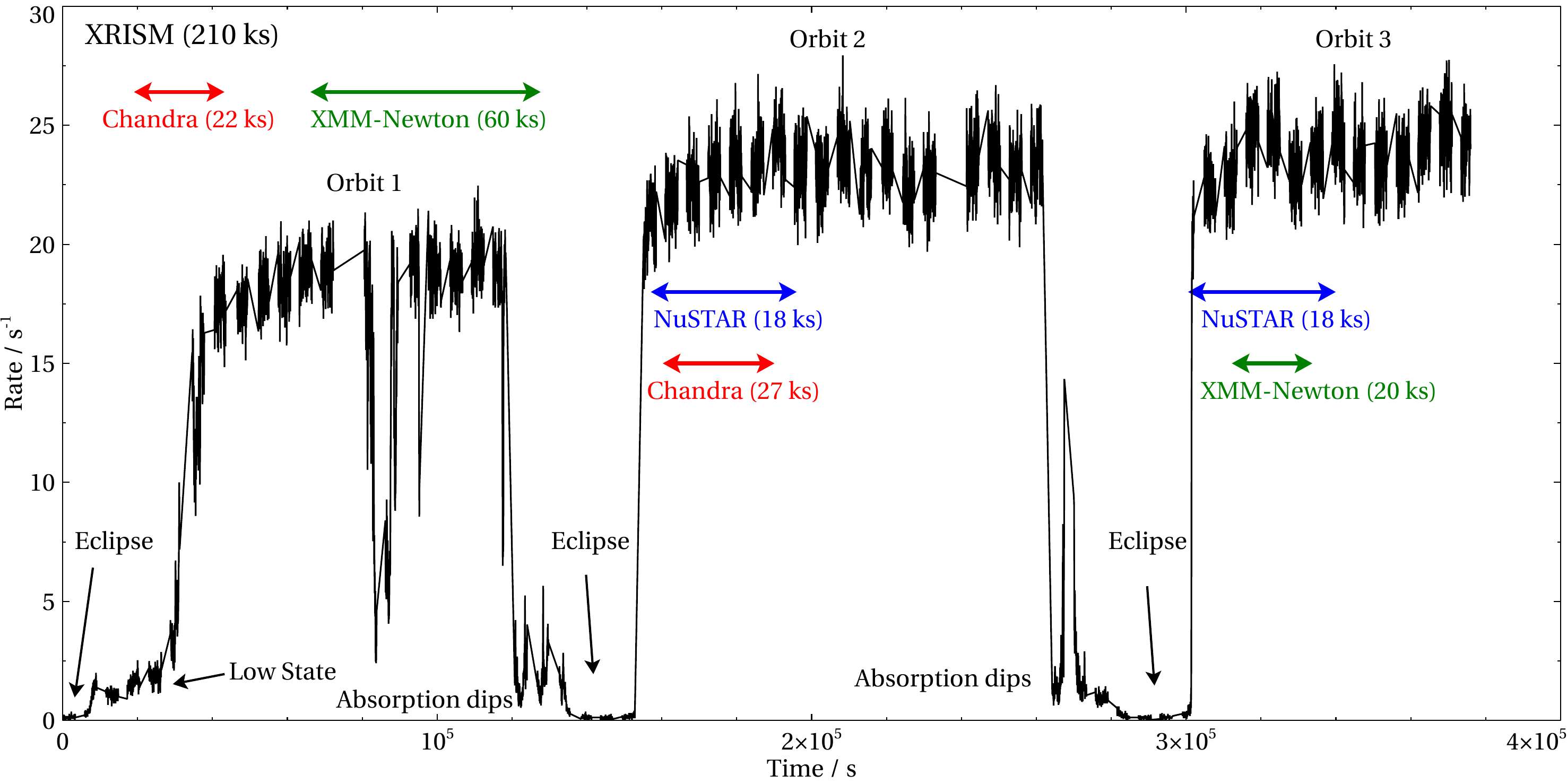}
\caption{Raw \xrism\ lightcurve (events of all qualities included) from the September 2024 campaign on Her X-1. Time T=0 corresponds to the beginning of the \xrism\ observation at MJD=60563.09653. Horizontal arrows show the overlaps of simultaneous observations with \xmm, \chandra\ and \nustar\ as well as their durations. The approximate clean exposures are given in the legend for each observatory. Some of the notable events are described, including our numbering scheme of individual Her X-1 orbits. \label{xrism_lightcurve}}
\end{figure*}

The \xrism\ Her X-1 observation began on September 10 2024 with a gross duration of 380 ks, resulting in a clean exposure of 210 ks. A raw \xrism\ lightcurve of the observation is shown in Fig. \ref{xrism_lightcurve}. The aim of the campaign was to begin observing as close as possible to the Turn-on of the Main High state, when our line of sight samples the disk wind at low heights above the disk (just after the disk itself uncovered our line of sight towards the X-ray source), and continue observing for a significant fraction of the High state as our sightline rises to greater heights above the disk. Thanks to an accurate Turn-on ephemeris prediction using the method of \citet{Staubert+16}, we were able to observe the transition from the Low into the Main High state roughly 30 ks after the beginning of the \xrism\ exposure. By using \xrism\ and \chandra\ lighcurves and finding the time when Her X-1 reached for the first time $\sim50$\% of the typical X-ray flux during the high-flux state, we determined that the Turn-on occurred at MJD=60563.457. We define the Her X-1 precession phase at time $t$ as $T_{0}+t/P$, where $T_{0}$ is the Turn-on time and $P$ is the precession cycle duration. The long-term average is 34.85 days, however for this specific cycle, the duration between the two consecutive Main High Turn-on moments was 34.04 days (Staubert et al. in prep.).

With \xrism\ we observed 3 nearly complete High state binary orbital periods \citep[the orbital period is roughly 1.7 days,][]{Tananbaum+72}. Hereafter, these orbits are denoted Orbit 1, Orbit 2 and Orbit 3, respectively. In addition to the high flux intervals, we also observed several periods of absorption dips and two complete and one partial eclipses. We note that this paper only focuses on high flux intervals, and all eclipse, Low state and absorption dip state data were excluded.

As the \xrism\ gate valve was closed, we also requested Director's Discretionary Time from \xmm\ \citep{Jansen+01} to capture the soft X-rays of Her X-1, primarily using the Reflection Grating Spectrometers (RGS). \xmm\ performed one observation which overlapped with Orbit 1, and one observation during Orbit 3 (80 ks total exposure time). We further triggered our Guaranteed Time observations with \chandra\ \citep{Weisskopf+00} and performed two simultaneous observations with the \chandra\ HETG gratings during Orbits 1 and 2, with a total exposure of 50 ks. Finally, to capture the hard X-rays above 17 keV, we triggered our Guest Observer program with \nustar\ \citep{Harrison+13} and observed Her X-1 for $\sim40$ ks in two snapshots during Orbits 2 and 3. The approximate timings of all simultaneous observations are shown alongside the \xrism\ lightcurve in Fig. \ref{xrism_lightcurve}. 

Further details of all observations taken during the 2024 campaign on Her X-1 are listed in Table \ref{obs_data_table} in Appendix \ref{app:data}.

\subsection{\xrism\ data reduction}

We reduced all \xrism\ data using \textsc{heasoft v6.35.1} with the \textsc{caldb} version 20250315. We followed standard data reduction routines described in the XRISM ABC Data Reduction Guide\footnote{https://heasarc.gsfc.nasa.gov/docs/xrism/analysis/}. We extracted events from the full Resolve array except the calibration pixel (12) and pixel 27, which is known to show abnormal gain behavior. We only used High and Medium quality primary events (Hp and Mp events), which were extracted into separate spectral files and fitted simultaneously without stacking. Typically, Mp event count rates were $10-20$\% of the Hp count rates during Her X-1 high flux periods analyzed in this paper. We generated the response matrix files and ancillary response files using \textsc{rslmkrmf} and \textsc{xaarfgen}, respectively, and used XL-sized \xrism\ response files for all spectral analysis. The data were binned according to the optimal binning scheme \citep{Kaastra+16} using the \textsc{ftool} routine \textsc{ftgrppha}. Finally, the data were converted from \textsc{ogip} format into \textsc{spex} format using the \textsc{trafo} routine. Since Her X-1 is a very bright source during its high flux periods, we did not use any generated \xrism\ background spectra. As the gate valve was closed during the Her X-1 observation, no source counts are registered below 1.8 keV. We therefore used the Resolve data between 1.8 and 12 keV. Data in the range between 12 and 17.5 keV were excluded as no atomic features are expected in this energy range and the higher statistics \nustar\ spectra are available. The \xrism\ data were separated into multiple intervals in order to perform a time-resolved wind study. Further details of the time-resolved \xrism\ data slices are listed in Appendix \ref{app:data}. We did not use any data from the Xtend instrument in this study, as its spectral resolution is not sufficient for a detailed disk wind analysis performed here.

\subsection{\xmm\ data reduction}

\xmm\ data were reduced using the \xmm\ Scientific Analysis System (SAS) package version 21.0.0. We processed the raw RGS \citep{denHerder+01} data with \textsc{rgsproc} using default source and background extraction regions. Considering the brightness of Her X-1, we did not exclude any periods of high background. The default background regions were strongly contaminated by source counts and hence we used blank field background spectra instead, following the same approach as \citet{Kosec+23a}. The final RGS spectral files were converted into \textsc{spex} format using the \textsc{trafo} routine, and binned by a factor of 3 directly within \textsc{spex} to achieve an oversampling of the instrumental resolution by roughly a factor of 3. RGS 1 and 2 data were always fitted simultaneously without any stacking. For spectral fitting, we used the 7.2 \AA\ (1.7 keV) to 36 \AA\ (0.35 keV) wavelength range.

We also used data from the European Photon Imaging Camera (EPIC) pn \citep{Struder+01}. The EPIC-pn camera was operated in Timing mode and the data were processed with the \textsc{epproc} routine and extracted using \textsc{evselect}. Only events of \textsc{pattern}~$\leq4$ (single and double) were accepted as valid events. No periods of background flaring had to be excluded for these observations. The source region was a rectangle centered on Her X-1 (RAWX coordinates between 15 and 56), and the background region was a narrow rectangle as far away on the chip from the source as possible (RAWX coordinates between 2 and 5). To mitigate pile-up, we also excluded a single central pixel column for observation 0953011401 and the central two pixel columns for observation 0953011501. The data were grouped using \textsc{specgroup} to oversample the spectral resolution of EPIC-pn by a factor of at most three and also to achieve Gaussian statistics, and finally converted into \textsc{spex} format using \textsc{trafo}. They were used in the energy range from 2 to 10 keV only for a comparison with the \xrism\ Resolve data and not for wind parameter estimates.

\subsection{\nustar\ data reduction}

The \nustar\ data were reduced using standard procedures with the NuSTAR Data Analysis Software. The data were first processed using the \textsc{nupipeline} routine, and the spectral files were extracted using the \textsc{nuproducts} routine. The source regions were circles centered on Her X-1 with a radius of 90 arcsec, and the background regions were larger circles away from the source with a radius of 140 arcsec. FPMA and FPMB data were extracted separately and fitted simultaneously without stacking. The spectra were binned with \textsc{ftgrppha} according to the optimal binning scheme and used in the energy range between 10 and 75 keV. The data below 10 keV were not needed as higher-resolution \xrism\ Resolve spectra were available. The $10-12$ keV region was used for cross-calibration between \xrism\ and \nustar.

\section{Results}
\label{sec:results}

All data were fitted in the \textsc{spex} fitting package \citep{Kaastra+96} version 3.07.03 by minimizing the Cash statistic \citep[C-stat,][]{Cash+79}. All uncertainties are provided at $1\sigma$ significance.

\begin{figure*}
\begin{center}
\includegraphics[width=0.7\textwidth]{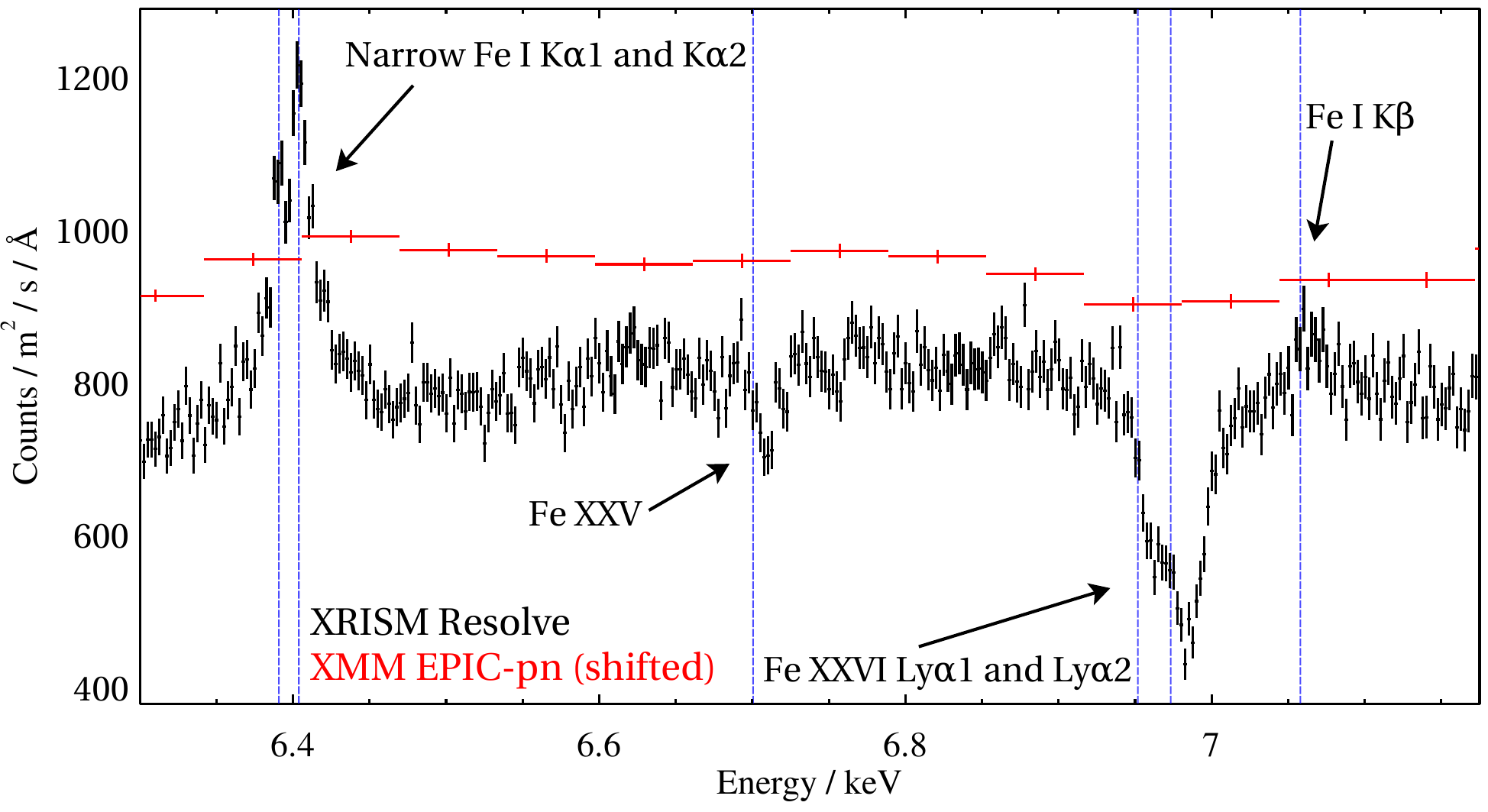}
\caption{Comparison between the time-averaged high-flux \xrism/Resolve spectrum, focusing on the Fe K energy band, with simultaneous data from \xmm\ EPIC-pn. The most notable spectral features are described and their rest-frame energies are shown with blue vertical dashed lines. The EPIC-pn data were shifted in energy to account for the known gain shift issue at high count rates \citep[see Appendix A of][]{Kosec+22}, and also in flux by a constant for visual purposes. \label{xrism_xmm_comparison}}
\end{center}
\end{figure*}

Thanks to the excellent spectral resolution of \xrism/Resolve, we are able for the first time to separate the various spectral features in the complex Fe K band of Her X-1. With previous \xmm\ and \chandra\ spectra, we were able to determine the presence of the strongest emission lines \citep{Kosec+22}, and the disk wind absorption \citep{Kosec+23a} in this band, but much uncertainty remained due to crowding of spectral features at these energies. The time-averaged high-flux \xrism/Resolve spectrum (from all 3 orbits combined), alongside a simultaneous \xmm\ EPIC-pn spectrum, is shown in Fig. \ref{xrism_xmm_comparison}.

\xrism\ clearly resolves several spectral components in the Fe K band. A narrow Fe I K$\alpha$ line is strongly detected and the K$\alpha$1 and K$\alpha$2 components are partially resolved. Blueshifted absorption from the accretion disk wind is also present. The strongest feature is the absorption line of Fe XXVI (Ly$\alpha$1 and Ly$\alpha$2 components are observed), but Fe XXV absorption is also present, although much weaker than Fe XXVI. The ratio of these two transitions indicates a high ionization degree of the wind. For the first time, we also detect a narrow Fe I K$\beta$ emission line. Finally, all of this narrow spectral structure is super-imposed on a highly broadened Fe K complex known from previous Her X-1 studies \citep{Kosec+22}.

All narrow components are broadened beyond the instrumental resolution of \xrism. This is expected considering the orbital motion of Her X-1 with a projected orbital velocity of 170 km/s \citep{Deeter+81}, which smooths all orbital-averaged spectra by this velocity width. Additionally, the narrow Fe I K$\alpha$ line is consistent with having at least 2 kinematic components, a very narrow one (with a width of $\sim100$ km/s) and a slightly broadened one (with a width of $\sim1000$ km/s).

\begin{figure}
\begin{center}
\includegraphics[width=\columnwidth]{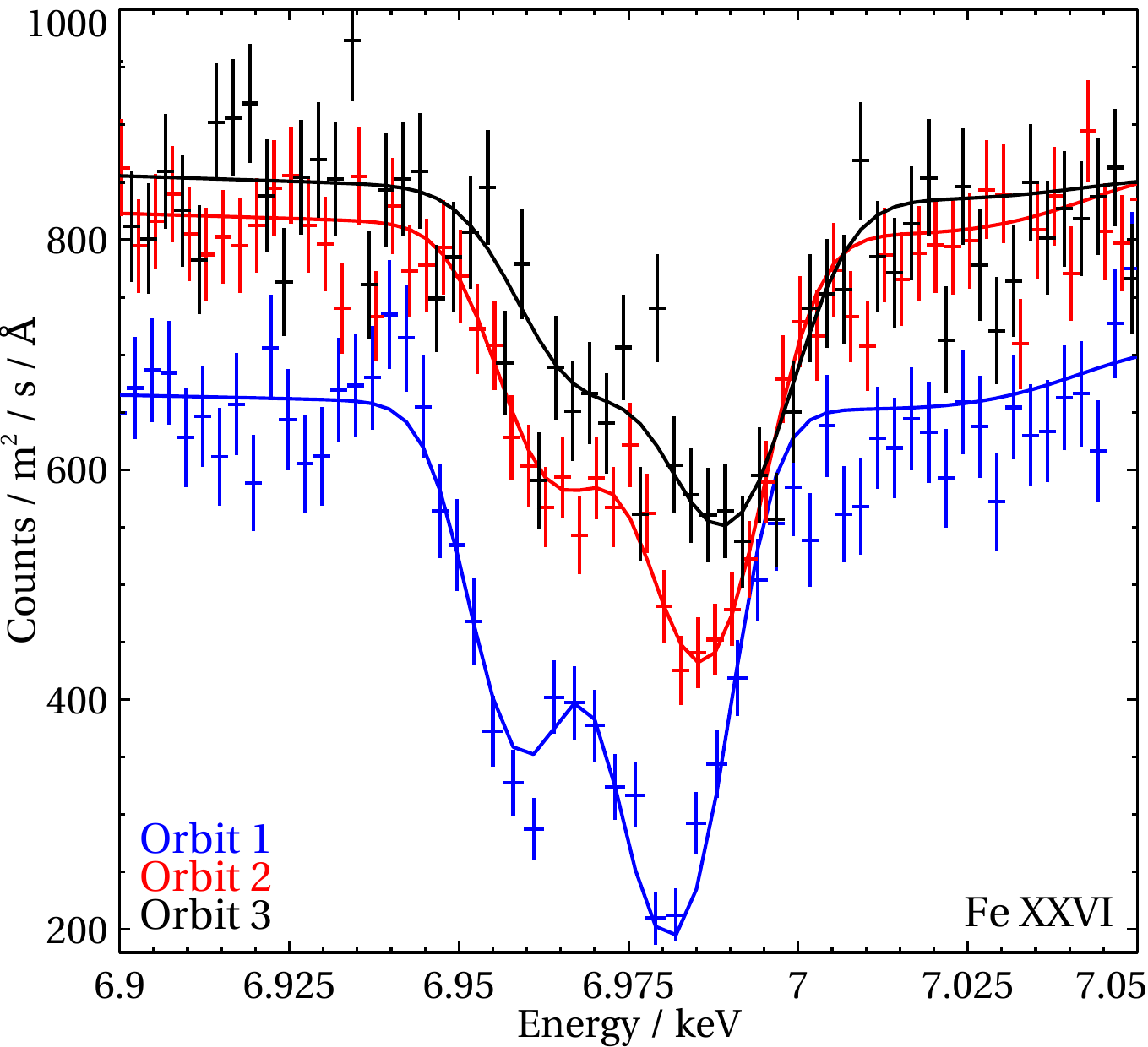}
\caption{Comparison of the Fe XXVI region between the 3 Her X-1 high-flux orbits. Fe XXVI absorption originates in the disk wind, and clear variability is observed from orbit to orbit. The line optical depth decreases, and it becomes broader and shifts to higher energies, indicating that the wind increases in velocity as well as velocity width. \label{FeXXVI_comparison}}
\end{center}
\end{figure}

Thanks to the very long exposure with \xrism, we are able to probe how the wind varies with time (and Her X-1 precession phase). To show this visually, we extracted data from the 3 Her X-1 orbits into separate spectra, and focused specifically on the Fe XXVI absorption line energy region. This is shown in Fig. \ref{FeXXVI_comparison}. The Fe XXVI doublet clearly varies from orbit to orbit. Its strength decreases over time from Orbit 1 to 2 to 3, as the 35-day cycle precession phase increases, in agreement with previous Her X-1 observations \citep{Kosec+23a}. For the first time, thanks to \xrism, we are also able to see shifts in its centroid energy and width over time. The line appears to broaden with time, and consistently shifts to higher energies. Therefore, the optical depth of the outflow in Fe XXVI decreases, and its projected velocity increases, as well as its velocity width. A similar evolution is observed also for Fe XXV, but at a lower signal-to-noise.

In the following sections we perform a more detailed time-resolved, quantitative analysis of these variations. The \xrism\ exposure is split into a number of intervals to understand how the relevant absorber parameters evolve with the phases of the orbital ($\sim1.7$ days) and super-orbital ($\sim35$ days) cycles seen in Her X-1. The interval sizes were defined adaptively to maximize the signal-to-noise within each interval, while preserving as much time resolution as possible, and are different for the different analyses performed below. The signal-to-noise depends on the combination of the X-ray flux seen from Her X-1 during the interval, both the optical depth and the width of the absorption lines, and the exact spectral model used to describe the absorber. We explored a range of interval sizes for each analysis, starting from small intervals and increasing their exposure until the best-fitting wind parameter uncertainties were small enough to track the systematic variation of the wind.

In subsection \ref{sec:modelsetup} we describe the setup of the spectral fitting model. In subsection \ref{sec:slab} we model the wind absorption features with a phenomenological spectral model which allows us to assess the absorption lines in the most model-independent manner. Finally, in subsection \ref{sec:pion} we apply a photoionization model to determine the physical parameters of the disk wind.

\subsection{Spectral model setup}
\label{sec:modelsetup}

We use a phenomenological spectral model to describe the continuum and line emission from Her X-1 before modeling the disk wind absorption. Each time interval of the time-resolved analysis is fitted independently. We fit the \xrism\ Hp and Mp data in two separate spectra using a cross-calibration constant. Its value is in all cases within 1\% of unity. The \xrism\ data are fitted simultaneously with \nustar\ data to obtain a coverage from 1.8 keV to 75 keV which encompasses most of the X-ray emission of Her X-1 (more than 90\% of its X-ray luminosity is in this range). The inclusion of the \nustar\ data is needed for correct spectral energy distribution (SED) modeling, necessary to apply a photoionization spectral model in subsection \ref{sec:pion}, and also to correctly describe the primary continuum across a broad band. Incorrect primary continuum modeling could have consequences for the inference of best-fitting parameters of the highly broadened Fe K emission lines as well as of the absorption lines. As we did not observe any significant variability in the count rate (which agrees within 5\%, Table \ref{obs_data_table}) and spectral shape between the two \nustar\ observations taken during Orbit 2 and Orbit 3, for each time interval analyzed we use the full \nustar\ spectrum from the appropriate observation. We introduce a cross-calibration constant between the \xrism\ and \nustar\ data to account for any calibration differences and to account for possible residual variability from interval to interval in the hard X-ray ($>12$ keV) flux during individual \xrism\ intervals. The value of the cross-calibration constant varies by less than 10\% for all analyzed intervals during Orbits 2 and 3 indicating weak, if any (pulse-averaged) variability in the hard X-rays. 

For Orbit 1 intervals, which lack simultaneous \nustar\ coverage (due to a spacecraft tracker visibility violation at that time period), we use the \nustar\ observation from Orbit 2. The cross-calibration constant for the analyzed time intervals during Orbit 1 is on average 30\% off the typical value during Orbits 2 and 3, indicating variability in the hard band. Nevertheless, inclusion of this (even if approximate) hard X-ray tail is much preferred to ignoring it completely. Assuming that the spectral shape in the hard X-ray band ($>12$ keV) remains consistent, the $\sim30\%$ flux variability is taken into account for SED modeling by applying the cross-calibration constant.

The primary continuum is described with a Comptonization component, using the \textsc{comt} model. This model was found to be insufficient alone to describe the entire $1.8-75$ keV energy band. A satisfactory fit was found when an additional hard exponential cutoff was applied using the \textsc{etau} component. The primary continuum is further affected by cyclotron resonance scattering at $\sim35-40$ keV \citep[one of the most famous spectral features of Her X-1,][]{Truemper+78}, which we reproduce using a \textsc{line} component. This is a strictly phenomenological description of the primary X-ray pulsar emission which allows us to focus on the interpretation of the spectral features in the Fe K band. 

\begin{figure*}
\begin{center}
\includegraphics[width=0.75\textwidth]{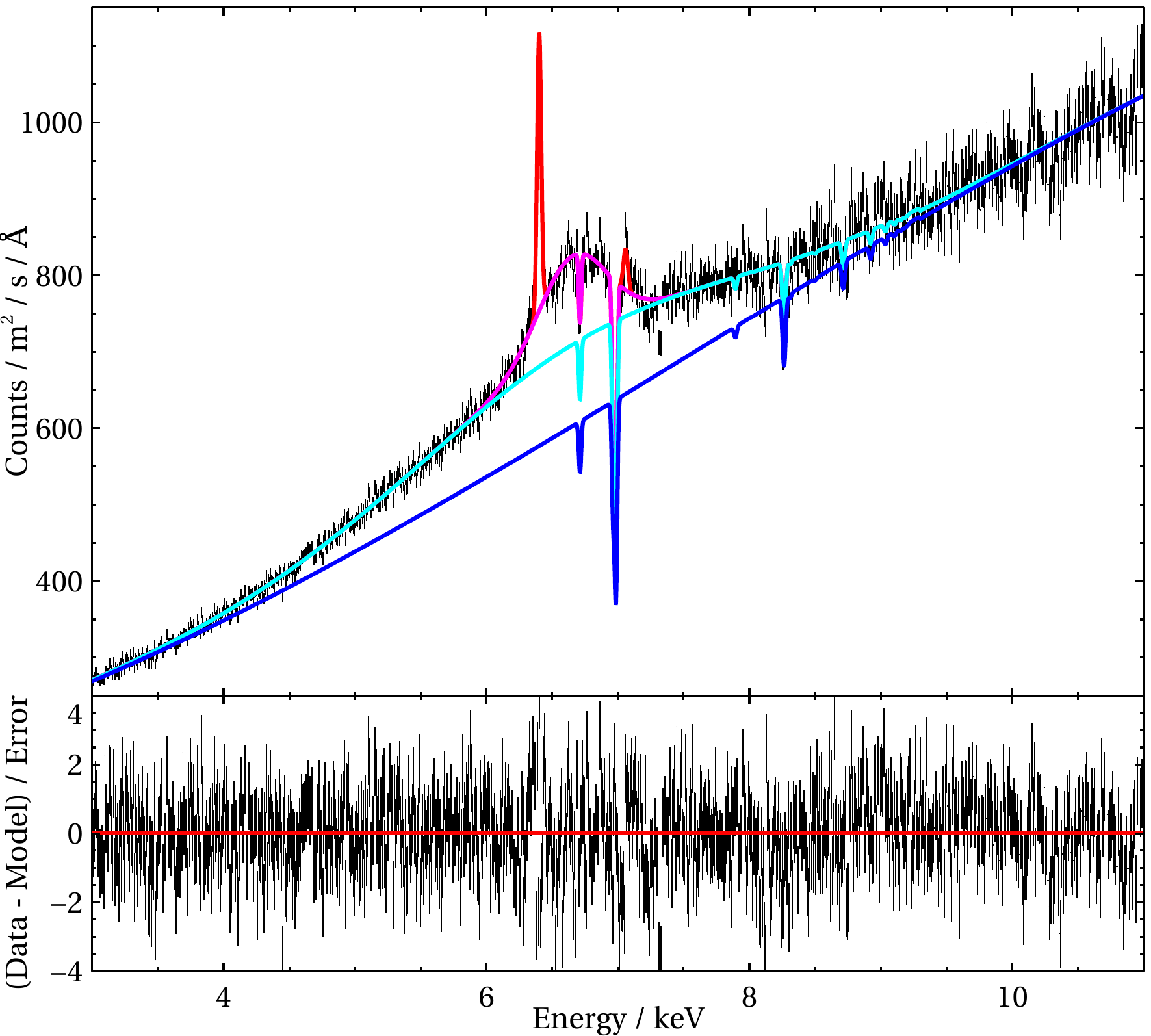}
\caption{Time-averaged high-flux \xrism\ spectrum of Her X-1 (all 3 orbits combined), fitted with the phenomenological emission and absorption spectral model. The primary continuum is shown in blue. Two highly broadened emission lines are required in addition to the primary continuum for a satisfactory fit of the Fe K complex (blue and magenta). Narrow Fe I K$\alpha$ and K$\beta$ emission lines are shown in red. On top of this continuum model, we apply disk wind absorption (using the \textsc{slab} model in this example) which produces the narrow absorption lines across the Fe K band. The lower panel contains the residuals to the fit containing all spectral components. \label{FeK_spectrum}}
\end{center}
\end{figure*}

We then focus on the Fe K band, which is shown in Fig. \ref{FeK_spectrum}. In agreement with our previous results from \xmm\ \citep{Kosec+22}, we require two highly broadened emission lines to accurately describe the broad Fe K structure on which the narrow emission and absorption lines are imprinted. These lines are modeled with simple Gaussians, which describe the spectral shape required by the \xrism\ data surprisingly well. One of the lines has a $1\sigma$ width of about 1 keV and its centroid is consistent with Fe I at 6.4 keV, while the second line has a $1\sigma$ width of $0.2-0.3$ keV and its centroid is instead consistent with the position of Fe XXV at 6.67 keV. The detailed properties of these features and their physical origin will be discussed in future work.

To describe the narrow Fe I K$\alpha$ and K$\beta$ lines, we also use single Gaussians. We note that we do not use 2 kinematic components to describe Fe I K$\alpha$ and K$\beta$ or the full Holzer line model \citep{Holzer+97} for these transitions in this work, as doing so would further increase the complexity of our model with very little effect on the absorber properties. As the origin of the two narrow lines is likely the same and the K$\beta$ line is relatively weak, we couple its velocity width to that of the K$\alpha$ component, and couple its energy to that of the K$\alpha$ component times the ratio of neutral K$\beta$ and K$\alpha$ rest-frame positions. The origin of the narrow Fe K$\alpha$ and K$\beta$ lines, and their multi-component kinematic structure will be studied in detail in Narita et al. (in prep.). Nevertheless, it is important to (at least crudely) model these features here for an accurate description of the emission continuum. Furthermore, the Fe I K$\beta$ component is located at $\sim7.07$ keV, close to the Fe XXVI transition at 6.96 keV, and so omitting the K$\beta$ model could have consequences for our description of the wind absorption through the Fe XXVI line. The time-averaged high-flux Her X-1 \xrism\ spectrum is shown in Fig. \ref{FeK_spectrum} and shows how the final spectral model consists of different model components.

Finally, the entire Her X-1 emission spectrum is subjected to Galactic absorption. This is included within the model with a \textsc{hot} component, which describes absorption by almost neutral gas \citep{dePlaa+04}. The column density of this absorber is set to $1\times10^{20}$ \pcm\ based on our previous study of Her X-1 \citep{Kosec+22, HI4PI+16}. This Galactic absorption model is applied as the last effect, only after the ionized disk wind absorber is applied.

\subsection{Phenomenological wind modeling}
\label{sec:slab}

The aim of this study is to probe how the disk wind varies with precession phase of Her X-1, as our sightline changes due to the warped disk precession. Therefore, we perform a detailed time-resolved study of the \xrism\ dataset. We chose a range of time-resolved intervals guided by the strength of the wind absorption lines and the statistics (and clean XRISM exposure) required to constrain them. As the first step, we perform phenomenological modeling of the wind absorption. We use the \textsc{slab} model \citep{Kaastra+02} in \textsc{spex}, which can determine the ionic column density in any specific ion by simultaneously fitting the optical depth of all the absorption lines which such ion produces. At the same time, the ratios of the different ion column densities are not constrained, and as such this spectral model gives us the most model-independent understanding of how the wind absorption lines behave over time.

The strongest features of the disk wind in the $2-10$ keV \xrism\ bandpass are the absorption lines of Fe XXV and Fe XXVI. In this first step, we simply use \textsc{slab} to fit for the column density of Fe XXV and XXVI ions. The \textsc{slab} model has two further variables - the absorber velocity (determined from the blueshift of the component) and the velocity width (determined from the line widths). This basic setup allows us to investigate any variation in the wind kinematics and in the strength of its wind absorption signatures.

\begin{figure}
\begin{center}
\includegraphics[width=\columnwidth]{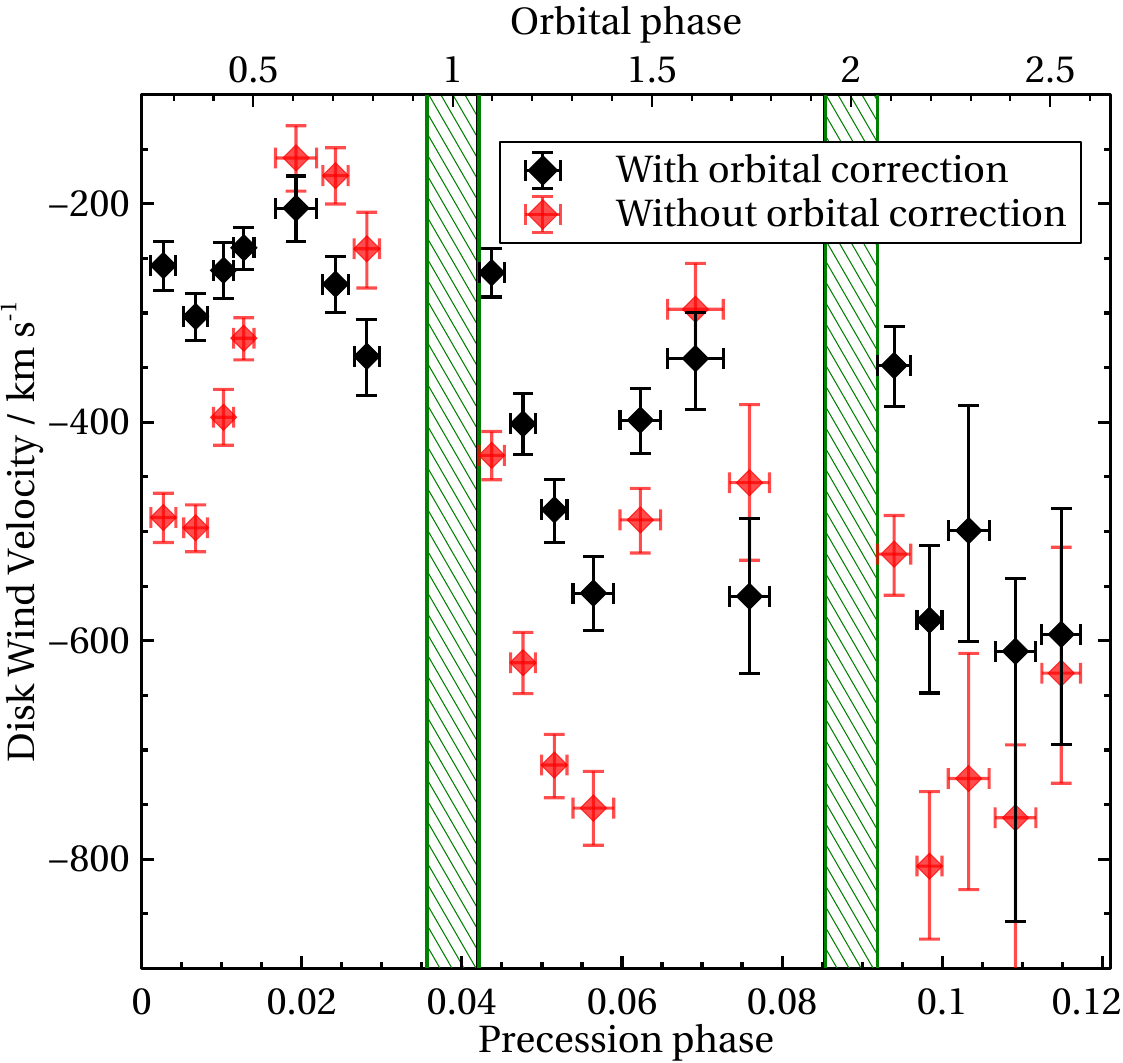}
\caption{Disk wind velocity versus Her X-1 precession phase and orbital phase. The best-fitting velocity with no corrections applied is in red. Clear orbital motion is observed in the evolution of this velocity. In black, we show the velocity evolution after correcting for the orbital motion of the neutron star. Her X-1 eclipses by the secondary star are shown with green horizontal lines. \label{Zv_evolution}}
\end{center}
\end{figure}

\xrism\ has the capability to track the kinematics of the wind of Her X-1 and its variations despite the low wind velocity of just hundreds of km/s. We observe that the wind velocity evolves in a range between 150 and 800 km/s. This is consistent with previous X-ray grating measurements in Her X-1 using \chandra\ HETG and \xmm\ RGS \citep{Kosec+20, Kosec+23a, Kosec+23c}. In fact, the \xrism\ data quality is so good that we can now observe binary orbital motion in the evolution of the wind velocity with precession phase. This is shown in Fig. \ref{Zv_evolution} by using the densest time-resolved sampling attempted in this study, where we split the high-flux \xrism\ exposure from all 3 orbits into 19 time intervals. Clear sinusoidal motion is observed on the timescale of the orbital period, which is expected because the wind is launched by the accretion disk orbiting the neutron star, and the neutron star is in orbital motion with the secondary star. On top of the orbital evolution, we also see systematic evolution in wind outflow velocity with precession phase. 

In the following analysis, we correct any wind outflow velocity measurements for the orbital motion of the neutron star and its accretion disk. We use the velocity correction formula from \citet{Kosec+20}, based on the measurement of the projected orbital velocity of $169.049 \pm 0.004$ km/s by \citet{Deeter+81} and the systemic velocity of Her X-1 of $-65 \pm 2$ km/s by \citet{Reynolds+97}. The orbital phase was calculated using the ephemeris of \citet{Staubert+16}, which specifies an orbital period of 1.700167590 days, an eclipse midpoint on MJD 46359.871940, and a change of the orbital period of $-(4.85 \pm 0.13) \times 10^{-11}$ s s$^{-1}$.

\begin{figure*}
\begin{center}
\includegraphics[width=0.85\textwidth]{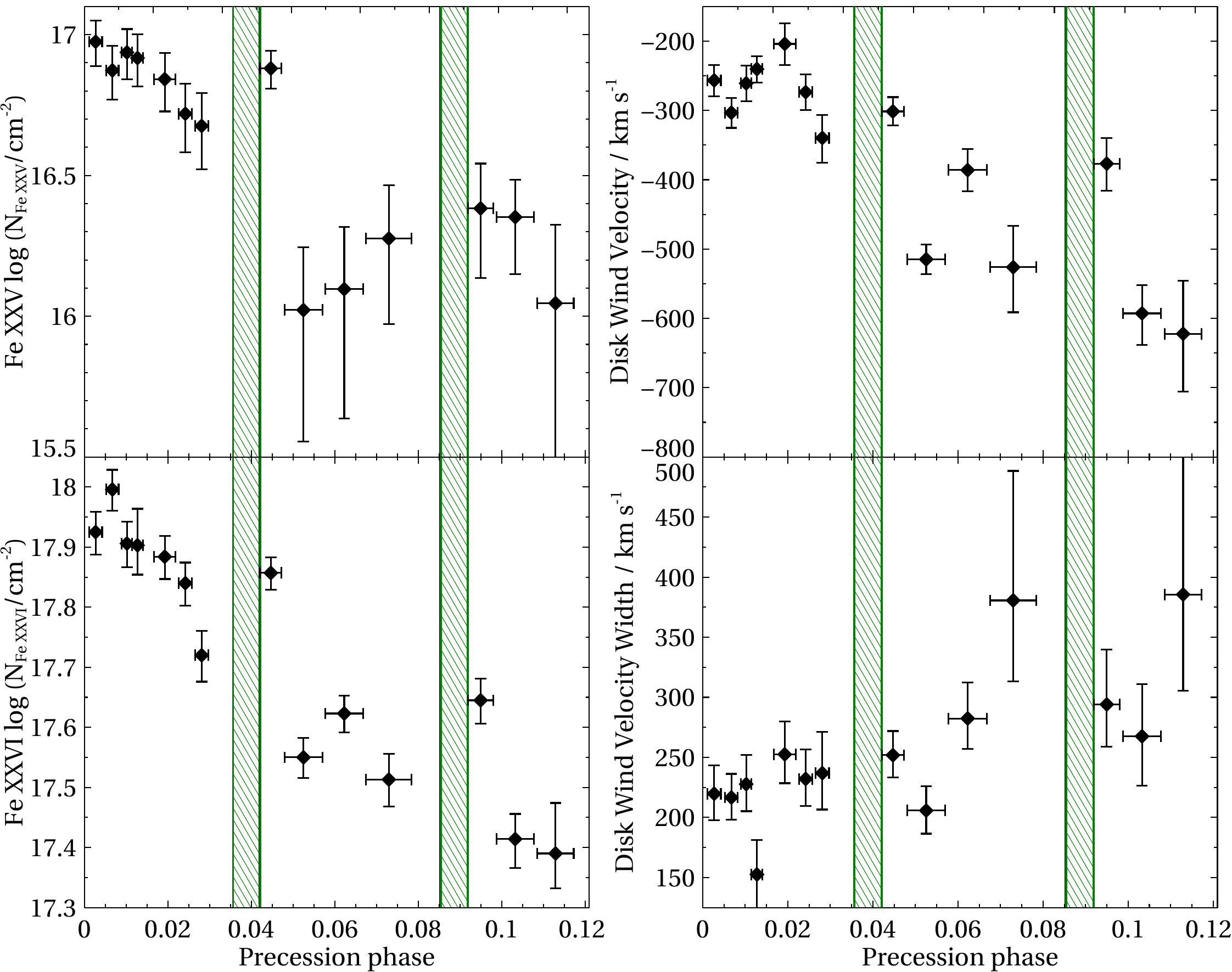}
\caption{Time-resolved phenomenological analysis of the disk wind properties using the \textsc{slab} model. The two panels on the left show the best-fitting ionic column densities of Fe XXV (top) and Fe XXVI (bottom). The two panels on the right show the kinematics of the wind (assuming the same kinematics for both lines), its outflow velocity (top) and the velocity width (bottom). The green vertical lines with shading indicate the Her X-1 eclipses.  \label{Slab_fig}}
\end{center}
\end{figure*}

We experimented with a range of time interval splitting which would maximize the time resolution while minimizing the uncertainties on the relevant parameters. In the end, a compromise was achieved by splitting the high-flux \xrism\ exposure into 14 intervals, where the exposure of each interval varied adaptively depending on the strength of the Fe XXV and XXVI wind absorption during each interval. The \textsc{slab} component free parameters were measured for each analyzed interval. The results are shown in Fig. \ref{Slab_fig} and listed in Table \ref{Slab_table}. We find that the wind is strongly variable, in agreement with previous studies and that all four \textsc{slab} parameters vary significantly over the sampled precession phase range. Additionally, the observed variations are not monotonic in most of the parameters, but show jumps just after Her X-1 emerges from eclipses by the secondary. We discuss these results in section \ref{sec:discussion}.

\begin{deluxetable}{cccccccc}
\tablecaption{Results of the phenomenological disk wind analysis using the \textsc{slab} model. \label{Slab_table}}
\tablewidth{0pt}
\tablehead{
\colhead{Orbit} & \colhead{Seg.} & \colhead{Precession} & \colhead{Fe XXV} & \colhead{Fe XXVI} & \colhead{Outflow} & \colhead{Velocity} & \colhead{C-stat/D.o.F.}\\
\colhead{} & \colhead{} & \colhead{phase} & \colhead{Column Density} & \colhead{Column Density} & \colhead{Velocity} & \colhead{Width} & \colhead{}\\
\colhead{} & \colhead{} & \colhead{} & \colhead{log(N$_{\rm Fe~XXV}$/\pcm)} & \colhead{log(N$_{\rm Fe~XXVI}$/\pcm)} & \colhead{km/s} & \colhead{km/s} & \colhead{}
}
\startdata 
1 & S1 & $ 0.00270 \pm 0.00156 $ & $ 16.98^{+0.075}_{-0.086} $ & $ 17.92^{+0.04}_{-0.05} $ & $ -260^{+20}_{-30} $ & $ 220^{+30}_{-20} $ & 5563.4/5474\\
& S2 & $   0.00673 \pm 0.00148 $ & $ 16.87^{+0.087}_{-0.10} $ & $ 18.00^{+0.04}_{-0.04} $ & $ -300^{+20}_{-20} $ & $ 220^{+20}_{-20} $  & 5862.4/5495 \\
& S3 & $   0.01022 \pm 0.00126 $ & $ 16.94^{+0.083}_{-0.096} $ & $ 17.91^{+0.04}_{-0.04} $ & $ -260^{+30}_{-30} $ & $ 230^{+30}_{-30} $ & 5824.6/5454\\ 
& S4 & $   0.01275 \pm 0.00128 $ &	  $ 16.92^{+0.084}_{-0.10} $ & $ 17.90^{+0.06}_{-0.05} $ & $ -240^{+20}_{-20} $ & $ 150^{+30}_{-40} $ & 5928.9/5488\\
& S5 & $   0.01928 \pm 0.00257 $ &	  $ 16.84^{+0.093}_{-0.11} $ & $ 17.88^{+0.04}_{-0.04} $ & $ -200^{+30}_{-30} $ & $ 250^{+30}_{-30} $ & 5729.6/5545\\
& S6 & $   0.02421 \pm 0.00158 $ &  $ 16.72^{+0.11}_{-0.14} $ & $ 17.84^{+0.04}_{-0.04} $ & $ -270^{+30}_{-30} $ & $ 230^{+30}_{-20} $ & 5698.2/5647\\
& S7 & $   0.02812 \pm 0.00158 $ &	  $ 16.68^{+0.12}_{-0.15} $ & $ 17.72^{+0.04}_{-0.05} $ & $ -340^{+40}_{-40} $ & $ 240^{+40}_{-30} $ & 5737.1/5622\\
\hline
2 & S8 & $ 0.04470 \pm 0.00258 $ &	  $ 16.88^{+0.063}_{-0.071} $ & $ 17.86^{+0.03}_{-0.03} $ & $ -300^{+20}_{-20} $ & $ 250^{+20}_{-20} $ & 6240.7/5943\\ 
& S9 & $   0.05253 \pm 0.00451 $ &	  $ 16.02^{+0.22}_{-0.47} $ & $ 17.55^{+0.03}_{-0.04} $ & $ -510^{+20}_{-20} $ & $ 210^{+20}_{-20} $ & 6228.4/6135\\
& S10 & $  0.06229 \pm 0.00451 $ & $ 16.10^{+0.22}_{-0.46} $ & $ 17.62^{+0.03}_{-0.03} $ & $ -390^{+30}_{-30} $ & $ 280^{+30}_{-30} $ & 6284.4/6115\\
& S11 & $  0.07299 \pm 0.00544 $ & $ 16.28^{+0.19}_{-0.30} $ & $ 17.51^{+0.05}_{-0.05} $ & $ -530^{+60}_{-70} $ & $ 380^{+110}_{-70} $& 6273.3/6131\\ 
\hline
3 & S12 & $ 0.09498 \pm 0.00307$ &	  $ 16.38^{+0.16}_{-0.25} $ & $ 17.65^{+0.04}_{-0.04} $ & $ -380^{+40}_{-40} $ & $ 290^{+50}_{-40} $ & 6176.6/5986\\
& S13 & $   0.10326 \pm 0.00447$ &	  $ 16.35^{+0.13}_{-0.20} $ & $ 17.41^{+0.04}_{-0.05} $ & $ -590^{+40}_{-50} $ & $ 270^{+50}_{-40} $ & 6228.6/6136\\
& S14 & $   0.11294 \pm 0.00435$ &	  $ 16.05^{+0.28}_{-0.81} $ & $ 17.39^{+0.09}_{-0.06} $ & $ -620^{+80}_{-90} $ & $ 390^{+260}_{-80} $ & 6500/6124\\
\enddata
\end{deluxetable}

\subsection{Photoionization wind modeling}
\label{sec:pion}

In the second part of the analysis, we replace the phenomenological \textsc{slab} model with the photoionization model \textsc{pion} \citep{Mehdipour+16} while keeping the other (continuum) spectral components the same. \textsc{pion} takes into account the SED of the currently loaded emission spectral model to calculate the ionization balance of all elements and produce their transmission function when subjected to the input SED, assuming photoionization equilibrium. Therefore, with \textsc{pion} we can take into account all of the wind absorption lines across the \xrism\ energy range, but at much higher computational cost. In return, we are able to determine physical parameters of the outflow such as its column density and ionization parameter, required to determine the outflow energetics and mass outflow rate. This is also naturally a more model-dependent approach to analysing the wind properties than using the phenomenological \textsc{slab} model which does not make assumptions on the conditions of the wind (e.g. single ionization phase versus multiphase), which is why we describe the results from both approaches in this manuscript. 

\textsc{pion} requires correct SED modeling to accurately determine the physical quantities. This is one of the main reasons why we add \nustar\ data to any \xrism\ time interval spectral fit. To also model the soft X-rays (0.35-1.8 keV) accurately, we perform a spectral fit of the RGS data from \xmm\, which was taken simultaneously with Orbits 1 and 3. For this step, the RGS data are used solely to determine the shape of the SED and not to constrain any of the wind absorption lines in the RGS band, considering we only have partial time coverage of the \xrism\ observations with \xmm. The RGS data are fitted with a simplified model from Appendix \ref{app:abundances} containing a simple blackbody (\textsc{bb}), a blackbody modified by coherent scattering (\textsc{mbb}) and one Gaussian line describing the broad 1 keV feature \citep[detailed reasoning for these specific model components is in][]{Kosec+22}. This simplified model accounts for more than 99\% of the soft X-ray flux. The model is added to each fitted \xrism\ time interval, and its best-fitting parameters (from the RGS analysis) are fixed. For all time intervals during Orbit 1, we use the results from the \xmm\ observation taken during that orbit. For all time intervals during Orbits 2 and 3, we use results from the \xmm\ observation taken concurrently with Orbit 3, considering that Orbits 2 and 3 show comparable count rates across the \xrism\ energy band (while Orbit 1 is somewhat fainter, Fig. \ref{xrism_lightcurve}). An example of the final, full-band SED (from Orbit 1) is shown in Fig. \ref{HerX1_SED}.

Additionally, \textsc{pion} assumes elemental abundances during the spectral fit. These could in principle be left to the default (Solar) values. However, there is strong evidence suggesting that elemental ratios in Her X-1 are in fact not Solar \citep{Jimenez+02, Jimenez+05, Kosec+23a, Kosec+23c}. Therefore, before proceeding with the time-resolved \textsc{pion} analysis, we perform the best-quality analysis of the wind elemental abundances given our simultaneous coverage with \xmm\ and \xrism\ that cover a wide energy band (0.35-12 keV), containing a broad range of elemental transitions. This elemental abundance analysis is described in further detail in Appendix \ref{app:abundances}. In the \textsc{pion} analysis described below, we fix the abundances to the values listed in Table \ref{el_abund}.

\begin{figure*}
\begin{center}
\includegraphics[width=0.85\textwidth]{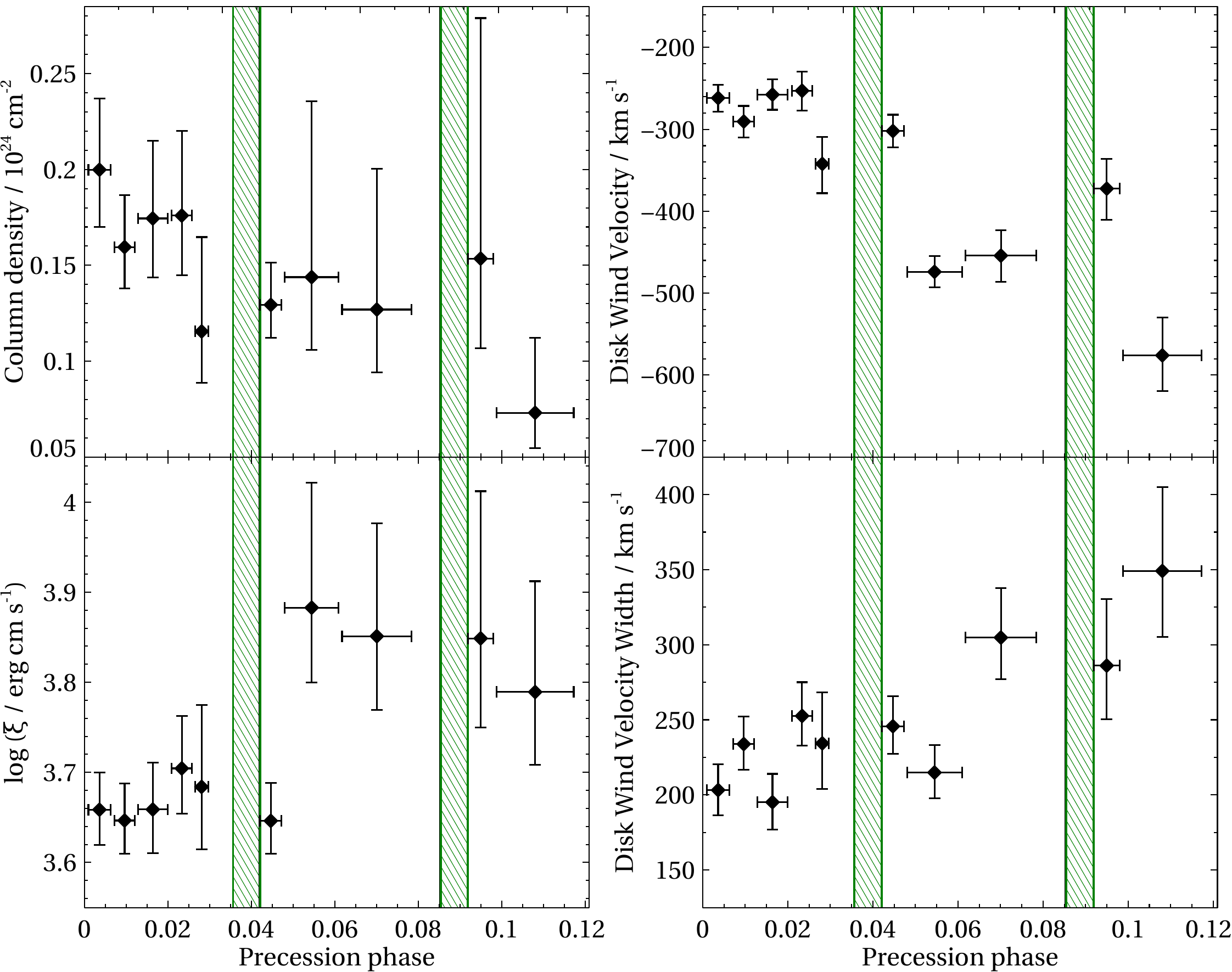}
\caption{Time-resolved physical photoionization analysis of the disk wind properties. The two panels on the left show the best-fitting disk wind column density N$_{\rm H}$ (top) and its ionization parameter \logxi\ (bottom), and the two panels on the right show the kinematics of the wind, its outflow velocity (top) and the velocity width (bottom). The green vertical lines with shading indicate the Her X-1 eclipses.  \label{Pion_fig}}
\end{center}
\end{figure*}

\begin{deluxetable}{ccccccccc}
\tablecaption{Results of the photoionization disk wind analysis using the \textsc{pion} model. \label{Pion_table}}
\tablewidth{0pt}
\tablehead{
\colhead{Orbit} & \colhead{Seg.} & \colhead{Precession phase} & \colhead{Column Density} & \colhead{\logxi} & \colhead{Outflow} & \colhead{Velocity} & \colhead{C-stat/D.o.F.} \\
\colhead{} & \colhead{} & \colhead{} & \colhead{} & \colhead{} & \colhead{Velocity} & \colhead{Width} & \colhead{} \\
\colhead{} & \colhead{} & \colhead{} & \colhead{$10^{24}$ \pcm} & \colhead{} & \colhead{km/s} & \colhead{km/s} & \colhead{}
}
\startdata 
1 & P1 & $ 0.00363 \pm 0.00264 $  & $ 0.200^{+0.037}_{-0.030} $ & $ 3.66^{+0.04}_{-0.04} $ & $ -262^{+16}_{-17} $ & $ 203^{+17}_{-17} $ & 6009.7/5707\\
& P2 & $   0.00966 \pm 0.00247 $  & $ 0.159^{+0.027}_{-0.022} $ & $ 3.65^{+0.04}_{-0.04} $ & $ -290^{+19}_{-20} $ & $ 234^{+18}_{-17} $ & 6137.6/5822\\
& P3 & $   0.01641 \pm 0.00354 $   & $ 0.174^{+0.041}_{-0.031} $ & $ 3.66^{+0.05}_{-0.05} $ & $ -258^{+19}_{-19} $ & $ 195^{+19}_{-18} $ & 5894.2/5618\\
& P4 & $   0.02336 \pm 0.00243 $   & $ 0.176^{+0.044}_{-0.031} $ & $ 3.70^{+0.06}_{-0.05} $ & $ -253^{+24}_{-24} $ & $ 253^{+22}_{-20} $ & 5946.1/5831\\
& P5 & $   0.02811 \pm 0.00157 $   & $ 0.116^{+0.049}_{-0.027} $ & $ 3.68^{+0.09}_{-0.07} $ & $ -342^{+33}_{-36} $ & $ 234^{+34}_{-30} $ & 5734/5622\\
\hline
2 & P6 & $ 0.04470 \pm 0.00258 $   & $ 0.129^{+0.022}_{-0.017} $ & $ 3.65^{+0.04}_{-0.04} $ & $ -302^{+20}_{-20} $ & $ 246^{+20}_{-19} $ & 6237.9/5943\\
& P7 & $   0.05449 \pm 0.00646 $    & $ 0.144^{+0.092}_{-0.038} $ & $ 3.88^{+0.14}_{-0.09} $ & $ -474^{+19}_{-19} $ & $ 215^{+18}_{-17} $  & 6419.5/6438\\
& P8 & $   0.07010 \pm 0.00833 $   & $ 0.127^{+0.073}_{-0.033} $ & $ 3.85^{+0.13}_{-0.08} $ & $ -454^{+31}_{-32} $ & $ 305^{+33}_{-28} $  & 6761.2/6573\\
\hline
3 & P9 & $ 0.09498 \pm 0.00307 $    & $ 0.153^{+0.130}_{-0.047} $ & $ 3.85^{+0.16}_{-0.10} $ & $ -372^{+36}_{-38} $ & $ 286^{+44}_{-36} $  & 6176/5986\\
& P10 & $  0.10804 \pm 0.00925 $    & $ 0.073^{+0.039}_{-0.018} $ & $ 3.79^{+0.12}_{-0.08} $ & $ -576^{+46}_{-43} $ & $ 349^{+56}_{-44} $  & 7061.7/6796\\
\enddata
\end{deluxetable}

The derived SEDs and elemental abundances are applied in a spectral fit to each of the time intervals independently. For this analysis, we had to reduce the number of time intervals to just 10. This was needed because \textsc{pion} requires multiple ionic transitions to accurately determine the column density and the ionization parameter - specifically, both Fe XXV and XXVI absorption lines are required, and the Fe XXV line is much weaker than the Fe XXVI line. Increasing the number of time intervals significantly expanded the uncertainty ranges on the column density and ionization parameter. The results of the \textsc{pion} analysis are shown in Fig \ref{Pion_fig} and listed in Table \ref{Pion_table}, and the individual segment spectra focusing on the Fe XXV and XXVI energy bands are shown in Fig. \ref{HerX1_combined_pion} in Appendix \ref{app:data}. As in the phenomenological analysis with the \textsc{slab} model, all parameters strongly vary with Her X-1 precession phase. The results are discussed in the following section.

\section{Discussion}
\label{sec:discussion}

We perform a comprehensive time-resolved analysis of the disk wind of Her X-1 by analyzing a long observation by \xrism. The total duration of $\sim380$ ks allows us to probe the wind properties for a significant fraction of the Main High state, spanning roughly 12\% of the entire 35-day precession period. The disk precession during this time shifts our line of sight across the wind vertical structure, resulting in a range of X-ray absorption spectra which give us a unique insight into the vertical properties of this structure. The observational campaign led by \xrism\ and supported by coordinated observations with \xmm, \nustar\ and \chandra\ also spanned almost 3 full high-flux (Main High state) orbital periods of Her X-1 and additionally captured 3 eclipses, 3 periods of absorption dips and a brief period of Low state. While this manuscript focuses on the measurement of time-resolved wind properties, our future publications will study many other aspects of this dataset including the pulse-resolved wind behavior, Her X-1 emission lines and other flux states.

For the first time, thanks to the spectral resolving power of \xrism\, we are able to completely resolve the complex Fe K band of Her X-1. \citet{Kosec+22} correctly identified several features in this energy band including the disk wind absorption lines (strongest being the Fe XXV and Fe XXVI lines), a narrow Fe K$\alpha$ line, a broadened Fe XXV line, and a highly broadened Fe K line. However, only now it is possible to measure the properties of these individual features with confidence. 

\subsection{Evolution of the disk wind with precession phase}

The time-resolved analysis shows that all parameters of the wind are strongly variable. We clearly observe the imprint of the Her X-1 binary orbital motion on the projected wind velocity (Fig. \ref{Zv_evolution}), showcasing the impressive performance of \xrism. Binary orbital motions were recently detected with \xrism/Resolve also in other X-ray binaries including Cen X-3 by \citet{Mochizuki+24} and in Cyg X-3 by\citet{XRISM+24} and \citet{Miura+25}. Even after correcting for the orbital motion effect, the Her X-1 projected outflow velocity systematically increases over time from 250 km/s to 600 km/s, as the precession phase increases and our line of sight samples greater heights of the wind above the warped disk. Neither the orbital motion, nor this systematic velocity evolution were previously detected with \xmm\ or \chandra\ \citep[see e.g. Extended Data Figure 4 of][]{Kosec+23a}.

The observation of an increasing velocity could indicate that the total wind velocity increases as the wind rises above the disk - in other words, the wind is still within the acceleration zone as it is sampled by our sightlines. Alternatively, at least its projected velocity must increase, i.e. as it rises higher above the disk, its velocity vector aligns more with our line of sight. In reality, both of these possibilities may be at play. The detection of an accelerating outflow would be of importance for determining the launching mechanism considering that thermally launched winds are ballistic, with no significant acceleration expected above the disk \citep{Begelman+83}. The detection of an accelerating motion would therefore indicate a magnetic origin of the wind, considering the relatively low Eddington fraction of Her X-1 ($0.1-0.2$) and the high ionization of the wind (likely excluding purely radiatively launched outflows). Therefore it is crucial to determine if the wind is indeed accelerating, or its streamlines are only being projected more along our line of sight as it rises higher above the disk.

In addition to the varying outflow velocity, we also observe a monotonic increase in the velocity width of the wind absorption lines from 200 to 400 km/s. This is unlikely to be due to an increase of thermal turbulent velocity (as the wind ionization parameter and temperature do not greatly increase over time), but instead is plausibly due to a range of streamlines crossing our line of sight, which all have a slightly different projected outflow velocity from our line of sight. If the velocity vectors of these streamlines diverge going upwards in the flow, the line velocity width will increase as our sightline moves to greater heights. Additionally, as each time interval has a definite exposure time, the Her X-1 orbital velocity change over this exposure duration will also contribute to the velocity width of the observed (time-integrated) wind features. However, this effect should not form the majority of the line velocity width, as the \textsc{slab} and \textsc{pion} analyses use different time interval samplings, and still show consistent velocity widths.

The column density of the wind, either in the individual lines (Fig. \ref{Slab_fig}) or in the overall N$_{\rm H}$ (Fig. \ref{Pion_fig}) is seen to decrease significantly over time. Particularly, the Fe XXVI transition appears to decrease almost linearly over time from $10^{18.0}$ cm$^{-2}$ down to $10^{17.4}$ cm$^{-2}$, with two exceptions (discussed below). However, the change in the overall N$_{\rm H}$ of the wind (from the \textsc{pion} analysis) from $2\times 10^{23}$ \pcm\ down to $7\times 10^{22}$ \pcm\  is not as sharp as in \citet{Kosec+23a} where a variation of as much as 3 orders of magnitude was found. Here, the change appears to be much lower, by about a factor of 3. This is because the behavior of the ionization parameter is very different compared with the previous analysis using \xmm\ data. Instead of strongly decreasing over time as seen in \citet{Kosec+23a}, \logxi\ is relatively stable in Orbit 1 at around a value of $3.65-3.7$, and increases in Orbit 2 and 3 to values consistent with $3.8-3.9$. There appears to be a step change in the ionization parameter after the first data point during Orbit 2 in the \textsc{pion} analysis (Fig. \ref{Pion_fig}), however it is likely that the best-fitting \logxi\ value of that data point is affected by possible presence of a multi-phase absorber structure along our line of sight during that time segment (discussed in the next subsection). The accretion disk wind \logxi\ variation may thus be more smooth than apparent from Fig. \ref{Pion_fig}, but in each case the variation of this parameter over the 2024 \xrism\ campaign is very limited in comparison with previous measurements.

One possibility is that the difference in the \logxi\ evolution between the \xrism\ and the older \xmm\ and \chandra\ observations is due to the great improvement in the dataset quality in the Fe K energy band thanks to the spectral resolution of Resolve (Fig. \ref{xrism_xmm_comparison}). Archival RGS observations \citep[e.g. Extended Data Figure 1 in][]{Kosec+23a} show a decrease of the strength of the N, O and Ne Ly$\alpha$ lines with precession phase. This can be explained either by a strong decrease in the column density, or an increase in the ionization parameter (or by a combination of both these variations). However, measuring the Ly$\alpha$ lines and their ratios in the RGS energy band is not sufficient to accurately measure the ionization of the wind. At these ionization parameters, \logxi\ is primarily constrained by the ratio of the Fe XXV and Fe XXVI transitions, which are poorly resolved in the archival \xmm\ EPIC data. The Fe XXV line is particularly weak, as can also be seen in the present \xrism\ study. The spectral fits based on EPIC data could then be affected by a degeneracy between the strength of the broadened Fe XXV emission line at 6.67 keV, and the optical depths of the Fe XXV and Fe XXVI wind absorption lines (see also Appendix \ref{app:abundances} for a further discussion of this effect). One supporting evidence for this hypothesis is that at low Her X-1 precession phases (phases $0-0.04$) - when the wind absorption lines are strong - the measurements by \xrism\ (Fig. \ref{Pion_fig}) are in agreement with those by \xmm\ and \chandra\ \citep[Fig. 2 in][]{Kosec+23a}. The measurements only significantly begin to diverge at greater precession phases (phase$>0.05$), when the wind features weaken.

An alternative possibility is that the wind structure has evolved since it was previously observed with \xmm\ and \chandra\ (during the large campaign in 2020 and before), and the differences between the parameter evolutions are real. Indeed, some change in the wind parameters is probably not unlikely. However, it is not clear if the evolution of the ionization parameter with precession phase could invert completely based on such changes. This would be of great interest as it would suggest that the disk wind of an X-ray binary can significantly evolve without an apparent change in any other relevant parameters such as the intrinsic source luminosity. In any case, both the current \xrism\ as well as the previous \xmm/\chandra\ observations agree that the wind absorption lines weaken with increasing precession phase.

Ultimately, now we can finally measure the ionization parameter with accuracy thanks to \xrism/Resolve. These measurements indicate (at least for this specific 35-day cycle) that as the wind rises to greater height above the disk, its ionization increases, and potentially stagnates. This is consistent with the wind `freezing out' at a constant \logxi\ as it rises and expands into 3D space. A \logxi\ stagnation would indicate that the wind number density evolves as $n \sim R^{-2}$ as it expands into space considering the ionization parameter definition of $\xi=L/(nR^{2})$, where L is the ionizing flux, $n$ is the wind number density and $R$ its distance from the ionization source. However, further \xrism\ observations at later Her X-1 precession phases would be necessary to confirm this potential trend, as well as definitively answer if the wind structure evolves over different 35-day cycles.

We note that the observed Her X-1 flux slightly increases over the \xrism\ observation, which can be seen in the increasing count rate in Fig. \ref{xrism_lightcurve}, by about 20\% between Orbit 1 and Orbit 2. However, this flux increase is insufficient to cause the observed increase in the value of ionization parameter. Considering the definition of \logxi\ (assuming a change in flux only, with no variation in number density and location), the Her X-1 flux would have to increase by 60\% to explain the observed \logxi\ variation by $\sim0.2$.

The different \logxi\ variation with precession phase inferred here has implications for the 2D map of the wind produced in our previous work \citep{Kosec+23a}. Qualitatively, this change in the ionization parameter evolution will push the wind location estimates (at certain phases/heights) with respect to the wind base closer to the neutron star. Hence, the wind streamline will be oriented more vertically than shown in Fig. 4 of \citet{Kosec+23a}.

We note that the significant uncertainties on the values of the column density are due to a degeneracy between \nh\ and \logxi. The uncertainties on \logxi\ are introduced because of the weakness of the Fe XXV transition, particularly in the second half of our campaign. This does not mean that the wind is detected weakly in these time segments, as it is still strongly detected through the Fe XXVI transition (Fig. \ref{Slab_fig}). In each of the individual segments, the statistical improvement upon adding the disk wind is at least \delcstat\ of 90, but typically reaches $200-500$.

We also use this combined X-ray dataset to measure the elemental abundances in the disk wind of Her X-1. This is discussed in further detail in Appendix \ref{app:abundances} and the best-fitting abundances are listed in Table \ref{el_abund}. In summary, the abundances of elements primarily in the RGS band (N and Ne) are over-abundant compared with O and consistent with previous studies \citep{Kosec+23a}. However, the Fe/O ratio is significantly different from the previously measured value of 2.3 due to our improved spectral resolution in the Fe K band and is $0.70_{-0.10}^{+0.08}$. 

In addition to modeling the wind ionized absorption, we also searched for re-emission from the regions of the outflow that are not along our line of sight. The detection of emitting wind plasma would help constrain its solid angle and the 3D wind structure in even more detail. This re-emission is very weak, even in the high-quality \xrism\ data. It is completely absent in the time-averaged high-flux spectrum (Fig. \ref{xrism_xmm_comparison}). There is evidence for weak re-emission in the orbit-resolved \xrism\ data, but only in Orbit 1 (Fig. \ref{FeXXVI_comparison}). This indicates that the outflow is being launched into a small solid angle, consistent with X-ray binary disk wind population studies \citep{Ponti+12, Parra+24}.

The large coordinated observational campaign on Her X-1 led by \xrism\ and supported by \xmm, \nustar\ and \chandra\ reveals the structure of its accretion disk wind in unprecedented detail. This first study of the 2024 dataset where we performed a time-resolved, pulse-averaged study of the wind shows how the structure changes upwards along the wind streamlines, as the precessing disk moves our sightline to consecutively greater heights above the disk. At the same time, the depth of this study only scratches the surface that the \xrism\ data provide, and opens many new questions discussed above. A follow-up study (Kosec et al. in prep) will explore the pulse-resolved variability of the wind properties, which has the power to constrain the wind number density via time-dependent photoionization modeling \citep{Kosec+24}, and precisely locate it along our line of sight to the neutron star. In future work, we will also produce an updated 2D map of the disk wind \citep{Kosec+23a}, taking into account the new \xrism\ measurements, and directly compare the wind parameter measurements and their evolution with precession phase (and height above the disk) with a range of hydrodynamical simulations assuming various disk wind driving mechanisms. Using this comparison, we will additionally address the question of the evolving wind velocity, to answer whether the wind is still located within the acceleration zone, or if only its projected velocity increases as it follows the streamlines and rises higher above the accretion disk.

The projected wind outflow velocity is seen to increase along the wind streamlines, from 250 km/s to 600 km/s. The question is - how much more can it increase further up the flow? If the wind is indeed thermally driven, a terminal velocity exceeding 1000 km/s would require very high Compton temperatures to accelerate the outflow to such speeds, or alternatively a contribution from radiation pressure would be needed. Future Her X-1 observations with \xrism\ at even later Her X-1 precession phases would help to answer this question and allow us to understand the wind properties even further along the streamlines, as the wind continues expanding into 3D space.

\subsection{Wind parameter variations with orbital phase - evidence for a second wind component?}

Most of the wind parameter evolution (after accounting for orbital motion) appears relatively smooth and monotonic, with two major exceptions. After each eclipse by the secondary, most wind parameters significantly jump. This is seen particularly in the Fe XXV and XXVI line column densities in Fig. \ref{Slab_fig}, which strongly increase for a single time segment. A similar shift is observed also in the outflow velocity (in both the \textsc{slab} and \textsc{pion} analyses). Surprisingly, no similar variability is seen in the velocity width, which remains consistent across the eclipses in both analyses. At the same time, no jumps in any wind parameters are seen just before the eclipses. However, we note that we lack coverage immediately preceding the eclipses, as these intervals are affected by pre-eclipse absorption dips (not analyzed in this study), which offer limited information on the wind properties (Fig. \ref{xrism_lightcurve}) given the much lower count statistics.

The observed jumps indicate a more complex nature of the disk wind than previously considered, showing behavior that was not seen in \xmm\ or \chandra\ observations of Her X-1. The jumps clearly show dependence on the orbital phase, and the large changes seem to only occur immediately following an eclipse, when our line of sight passes close to the secondary star. This may indicate multiphase nature of the wind, with one component originating from the Her X-1 accretion disk and following a smooth evolution (seen particularly in the Fe XXVI line column density), and a second component, plausibly originating from the secondary star (considering the orbital phase dependence) and only appearing over a specific range of orbital phases. We note that if there are indeed two absorber components, our \textsc{pion} analysis may be affected in the two data points immediately following the eclipses as it strictly assumes one component with a single ionization parameter \logxi. If two components are present with different ionization parameters, the best-fitting single \textsc{pion} model parameters will be biased. This would explain why the \textsc{slab} ionic column density jumps for both Fe XXV and XXVI after the first eclipse but the same does not occur for the overall \textsc{pion} column density. 

In the case of two wind components with slightly different projected outflow velocities causing the jumps in the absorber systematic velocity, we would in principle also expect an increase in the absorber velocity width. This is surprisingly not observed, but it could again be an artifact of the fitting process driven by the usage of a single kinematic model, or a limitation of the dataset even with the resolution power of \xrism/Resolve.

One of the possible launching mechanisms for this outflow, if it is launched from the secondary star, may be the strong X-ray illumination of the secondary by Her X-1, which could launch a thermal wind from the stellar surface. The existence of such a wind was previously hypothesized by \citet{Boroson+01} based on the detection of P-Cygni profiles in the UV band. However, we note that these UV line detections in Her X-1 predate most X-ray binary disk wind confirmations (in the X-rays) and particularly the disk wind detection in this source \citep{Kosec+20}. It is thus possible that the UV lines are imprinted instead by the accretion disk wind. A re-analysis of the original HST observations from \citet{Boroson+01} could inform us of their relation to the absorber observed in the X-ray band.

An alternative possibility is that there is a single disk wind component with its streamlines focused by the secondary star's gravity, the effect of which is only observable close to the eclipses. However, the previous estimates of the wind's distance from the X-ray source all place the wind location within the outer accretion disk radius, far away from the stellar influence. A more complex spectral analysis containing multiple absorption zones is required. This will be explored in follow-up work.

\section{Conclusions}
\label{sec:conclusions}

We present the first results from a large coordinated campaign on Hercules X-1 led by \xrism\ (210 ks exposure) and supported by \xmm\ (80 ks), \nustar\ (40 ks) and \chandra\ (50 ks). This work focuses on the properties of the vertical structure of the disk wind of Her X-1, which we measured by performing a time-resolved analysis of the observations. We applied both phenomenological as well as physical spectral models to describe the ionized absorption imprinted by the outflow. Our conclusions are as follows:

\begin{itemize}

    \item For the first time, thanks to the excellent spectral resolution of \xrism, we are able to fully resolve the complex Fe K energy band of Her X-1, and accurately measure the properties of the individual components. In particular, the absorption lines of Fe XXV and Fe XXVI from the accretion disk wind are resolved and allow us to accurately measure the ionization parameter of the wind.
 
    \item We directly observe the orbital motion of Her X-1 in the evolution of the wind outflow velocity. After correcting for this motion, the wind velocity is seen to increase with rising precession phase from 250 km/s to 600 km/s, as our sightline progressively samples the wind streamlines at greater heights above the disk. This indicates that either the wind is still within the acceleration zone at these heights, or that its velocity vector progressively aligns with our line of sight with increasing height above the disk. Additionally, we also observe a rise in the wind velocity width with increasing precession phase, from 200 to 400 km/s.
    
    \item The column density of the wind decreases with increasing height above the disk, in agreement with previous observations, from $20\times10^{22}$ \pcm\ to $7\times10^{22}$ \pcm. This decrease is also observed in the individual Fe XXV and XXVI absorption lines. In contrast to previous results, the ionization parameter \logxi\ increases with the precession phase, from 3.65 to $3.8-3.9$ and stagnates around this value as the wind expands into 3D space.

    \item The wind parameter evolution is relatively smooth throughout the observation, except immediately after each eclipse, when a significant increase in ionized absorption is observed. We also detect a jump in the wind velocity at these phases. This suggests that the ionized absorption is multiphase, and a second component appears only briefly after eclipses. The second component may originate from the secondary star and could be driven thermally by the strong illumination of the secondary by the intense X-ray radiation from Her X-1.

    \item The improved spectral data quality allows us to perform a thorough study of elemental abundances within the disk wind. We are able to constrain the abundances of C, N, O, Ne, Mg, Si, S, Ar, Ca, Fe and Ni. Their best-fitting values are listed in Table \ref{el_abund}.

\end{itemize}

\begin{acknowledgments}
Support for this work was provided by NASA through the NASA Hubble Fellowship grant HST-HF2-51534.001-A awarded by the Space Telescope Science Institute, which is operated by the Association of Universities for Research in Astronomy, Incorporated, under NASA contract NAS5-26555. PK also aknowledges support from NASA grants 80NSSC25K7317 and 80NSSC25K7533. CP is supported by European Union - Next Generation EU, Mission 4 Component 1 CUP C53D23001330006. DJW acknowledges support from the Science and Technology Facilities Council (STFC; grant code ST/Y001060/1). This paper employs a list of Chandra datasets, obtained by the Chandra X-ray Observatory, contained in the Chandra Data Collection ~\dataset[DOI: 10.25574/cdc.569]{https://doi.org/10.25574/cdc.569}.
\end{acknowledgments}





%
\facilities{\xrism, \xmm, \nustar, \chandra}

\software{Veusz, SPEX \citep{Kaastra+96}, XSPEC \citep{Arnaud+96}
          }


\appendix

\section{Further details of individual observations and time-resolved data intervals}
\label{app:data}

\begin{figure*}
\begin{center}
\includegraphics[width=\textwidth]{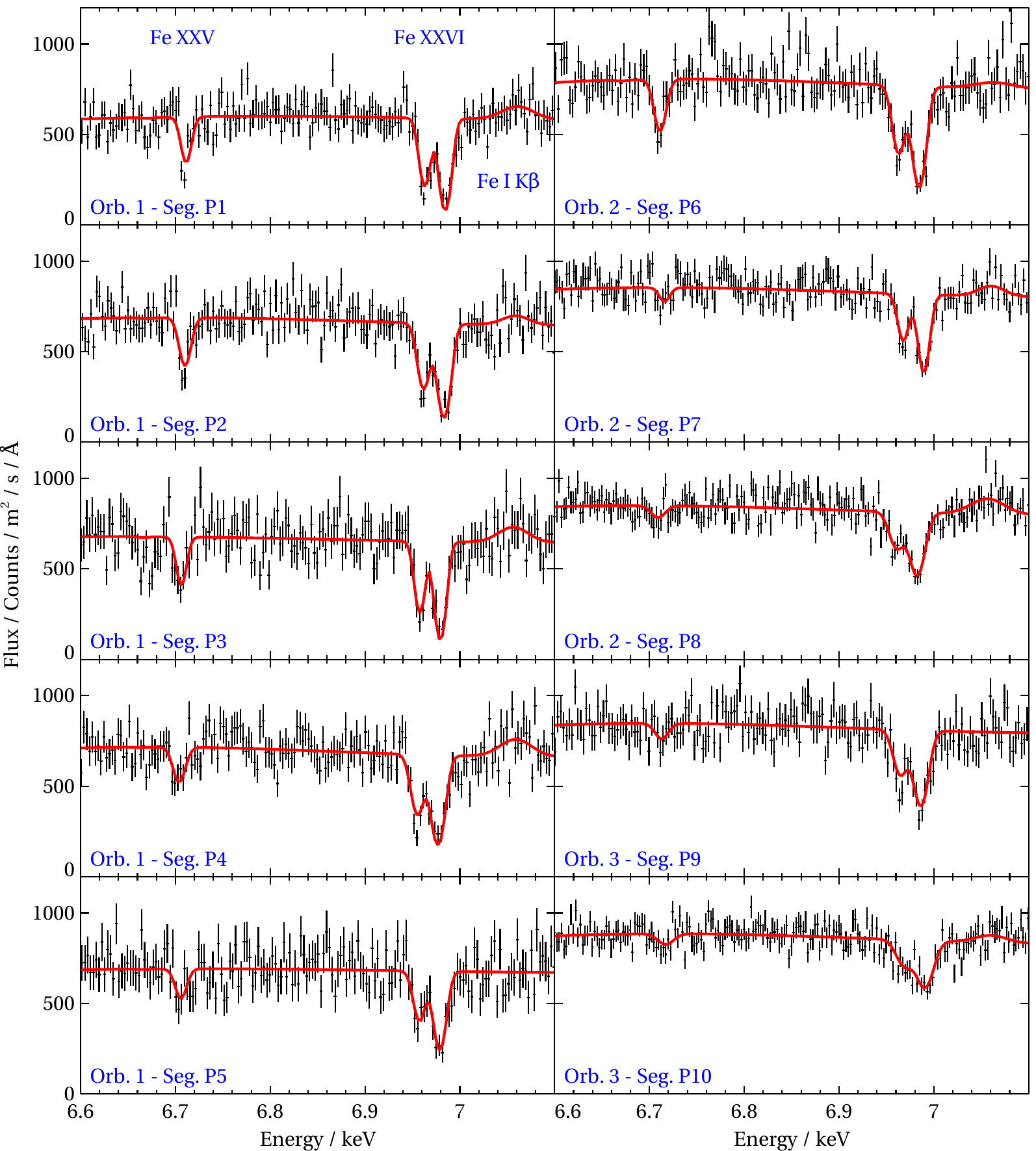}
\caption{Narrow-band \xrism\ spectra  focusing on the Fe XXV and XXVI energy region from the physical time-resolved analysis (using the \textsc{pion} model) split into 10 segments alongside with the best-fitting spectral model. Only Hp event spectra are shown. The top left panel includes labels describing the different spectral features observed in Her X-1. We use the segment numbering scheme (from P1 to P10) defined in Table \ref{Pion_table}. The X and Y axes ranges are the same for all panels. \label{HerX1_combined_pion}}
\end{center}
\end{figure*}

Table \ref{obs_data_table} contains details about the all coordinated observations during the 2024 campaign on Her X-1. We explored a range of segmenting approaches by splitting the \xrism\ dataset into 10 intervals (used in the \textsc{pion} analysis in section \ref{sec:pion}), 14 intervals (used in \textsc{slab} analysis in section \ref{sec:slab}) and 19 intervals (used for the velocity evolution Fig. \ref{Zv_evolution}). The details in Table \ref{xrism_slab_data_table} are given for the 14 segment approach and in Table \ref{xrism_pion_data_table} for the 10 segment approach. Both tables include the clean exposure and the mean Hp and Mp quality event count rate of each segment as well as the segment center points in MJD and in Her X-1 precession phase. Fig. \ref{HerX1_combined_pion} contains the narrow-band \xrism\ spectra from the 10-segment analysis (focusing on the Fe XXV and XXVI energy region) including the best-fitting \textsc{pion} spectral model.

\begin{table*}
\caption{Details of all observations during the 2024 coordinated campaign on Her X-1. The exposure value is the clean exposure from all intervals of high flux after excluding all periods of time during which Her X-1 was exhibiting Low state, eclipses and absorption dips.\label{obs_data_table}}
\begin{tabular}{ccccccc}
\hline
\hline
Observatory & Observation ID & Start Time & End Time & Instrument & Exposure & Count Rate\\
 &  & & &  & s &  s$^{-1}$\\
\hline
\xrism&201074010&2024-09-10 02:19&2024-09-14 10:46&Resolve&144451&16.7\\
\nustar&81002350002&2024-09-11 21:51&2024-09-12 08:51&FPMA&17953& 62.4\\
&&&&FPMB&18024&58.5\\
\nustar&81002350004&2024-09-13 13:51&2024-09-14 00:56&FPMA&18120&63.2\\
&&&&FPMB&18192&59.0\\
\xmm&0953011401&2024-09-10 20:55&2024-09-11 13:34&RGS 1&37552&8.64\\
&&&&RGS 2&37290&9.56\\
\xmm&0953011501&2024-09-13 17:18&2024-09-13 22:52&RGS 1&18270&11.0\\
&&&&RGS 2&18162 &12.1\\
\chandra&28768&2024-09-10 07:53&2024-09-10 09:33&MEG&5980.8&5.02\\
&&&&HEG&5980.8&3.47\\
\chandra&29528&	2024-09-11 23:08&2024-09-12 07:24&MEG&27518&7.11\\
&&&&HEG&27518&4.91\\
\hline
\end{tabular}
\end{table*}

\begin{table}
\caption{Details of time-resolved \xrism\ analysis, where all high-flux periods of time were split into 14 segments, used in the phenomenological \textsc{slab} wind analysis in section \ref{sec:slab}. The values of MJD and precession phase refer to the centroids of the relevant segments. The uncertainties on the MJD and precession phase indicate the segment duration rather than statistical uncertainties. \label{xrism_slab_data_table}}
\begin{tabular}{ccccccc}
\hline
\hline
Orbit & Seg. & MJD & Precession phase & Clean Exposure & Hp Count Rate& Mp Count Rate \\
 &  &  &  & s &  s$^{-1}$ & s$^{-1}$ \\
\hline
1 & S1 & $ 60563.549 \pm 0.053 $ & $ 0.00270 \pm 0.00156 $ &6475 &12.98 &1.35\\
& S2 & $ 60563.686 \pm 0.050 $ & $   0.00673 \pm 0.00148 $ & 5724& 14.53&1.66\\
& S3 & $ 60563.804 \pm 0.043 $ & $   0.01022 \pm 0.00126 $ &5108 &14.81 &1.83\\
& S4 & $ 60563.891 \pm 0.043 $ & $   0.01275 \pm 0.00128 $ &5209 &15.25 &1.86\\
& S5 & $ 60564.113 \pm 0.087 $ & $   0.01928 \pm 0.00257 $ &5922 & 14.92&1.81\\
& S6 & $ 60564.281 \pm 0.054 $ & $   0.02421 \pm 0.00158 $ &6899 &15.23 &1.85\\
& S7 & $ 60564.414 \pm 0.054 $ & $   0.02812 \pm 0.00158 $ &6611 &15.22 &1.88\\
\hline
2 & S8 & $ 60564.978 \pm 0.088 $ & $ 0.04470 \pm 0.00258 $ &10325 &16.61 &2.28\\
& S9 & $ 60565.245 \pm 0.153 $ & $   0.05253 \pm 0.00451 $ &17010 &17.51 &2.57\\
& S10 & $ 60565.577 \pm 0.153 $ & $  0.06229 \pm 0.00451 $ &16176 &17.28 &2.53\\
& S11 & $ 60565.941 \pm 0.185 $ & $  0.07299 \pm 0.00544 $ &16981&17.46 &2.55\\
\hline
3 & S12 & $ 60566.690 \pm 0.104 $ & $ 0.09498 \pm 0.00307 $ &10712 &17.31 &2.49\\
& S13 & $ 60566.972 \pm 0.152 $ & $   0.10326 \pm 0.00447 $ &16974 &17.73 &2.62\\
& S14 & $ 60567.301 \pm 0.148 $ & $   0.11294 \pm 0.00435 $ &16142 &17.86 &2.66\\
\hline
\end{tabular}
\end{table}

\begin{table}
\caption{Details of time-resolved \xrism\ analysis, where all high-flux periods of time were split into 10 segments, used in the physical \textsc{pion} wind analysis in section \ref{sec:pion}. The values of MJD and precession phase refer to the centroids of the relevant segments. The uncertainties on the MJD and precession phase indicate the segment duration rather than statistical uncertainties. \label{xrism_pion_data_table}}
\begin{tabular}{ccccccc}
\hline
\hline
Orbit & Seg. & MJD & Precession phase & Clean Exposure & Hp Count Rate& Mp Count Rate \\
 &  &  &  & s &  s$^{-1}$ & s$^{-1}$ \\
\hline
1& P1 & $ 60563.580 \pm 0.090 $ & $ 0.00363 \pm 0.00264 $ &9393 & 13.39&1.44\\
& P2 & $ 60563.785 \pm 0.084 $ & $  0.00966 \pm 0.00247 $ & 9776& 14.87&1.80\\
& P3 & $ 60564.015 \pm 0.120 $ & $  0.01641 \pm 0.00354 $ &6804 &14.92 &1.79\\
& P4 & $ 60564.252 \pm 0.083 $ & $  0.02336 \pm 0.00243 $ & 9413&15.26 &1.88\\
& P5 & $ 60564.413 \pm 0.054 $ & $  0.02811 \pm 0.00157 $ &6611 &15.22 &1.88\\
\hline
2& P6 & $ 60564.978 \pm 0.088 $ & $ 0.04470 \pm 0.00258 $ &10325 &16.61 &2.28\\
& P7 & $ 60565.311 \pm 0.220 $ & $  0.05449 \pm 0.00646 $ &23864 &17.48 &2.57\\
& P8 & $ 60565.843 \pm 0.284 $ & $  0.07010 \pm 0.00833 $ &26303 & 17.37&2.54\\
\hline
3& P9 & $ 60566.690 \pm 0.104 $ & $ 0.09498 \pm 0.00307 $ &10712 &17.31 &2.49\\
& P10 & $ 60567.134 \pm 0.315 $ & $ 0.10804 \pm 0.00925 $ &33147 &17.79 &2.64\\
\hline
\end{tabular}
\end{table}

\section{Elemental abundance analysis}
\label{app:abundances}

We leverage the simultaneous \xrism, \xmm\ and \nustar\ data to perform the most detailed elemental abundance analysis of the disk wind of Her X-1 to date. We previously analysed the wind abundances in \citet{Kosec+23a} and \citet{Kosec+23c} using \xmm\ and \chandra\ data separately. However, the spectral resolution of \xrism\ in the Fe K band allows us to perform a much more robust analysis by resolving the complex structure in this energy band, and accurately measuring the wind ionization parameter. \xmm\ RGS data inform us about the abundances of lighter elements such as C, N, O, Ne and Mg with transitions in the $0.35-1.75$ keV band. On the other hand, the energy band of \xrism\ ($1.8-12$ keV) contains transitions of heavier elements including Si, S, Ar, Ca, Fe and Ni. Therefore, to extract the maximum amount of information, we combine both datasets. Since the disk wind is highly ionized and its signatures are the metallic lines of these elements (with no H lines within this X-ray range), we cannot extract the elemental abundances with respect to the abundance of hydrogen. Instead, we must compare the abundance with respect to one reference metal. Following our previous work \citep{Kosec+23a}, we primarily choose O to be the reference, and fix its abundance to 1. This choice is used in the \textsc{pion} spectral analysis in section \ref{sec:pion}. For completeness, we also list the best-fitting abundance values assuming Fe as the reference element. This choice allows a direct comparison with the abundance study of disk winds in 4 low-mass X-ray binaries by \citet{Keshet+25}.

To leverage as much information as possible, we use all data from the 2024 campaign on Her X-1. However, as the disk wind is strongly variable, the data must be split and care must be taken in any simultaneous fitting. The \xmm\ observations occurred during \xrism\ Orbit 1 and 3 but covered only parts of those orbits (compared with \xrism, which covered full orbits). For this reason, we split the \xrism\ data from those orbits into two spectral data groups - one spectrum simultaneous with the \xmm\ pointing, and one spectrum non-simultaneous with \xmm. There was no \xmm\ observation during Orbit 2, and so those \xrism\ data form their own data group. We also add the \nustar\ observations in each data group for accurate SED modeling (required by the \textsc{pion} wind model) - the \nustar\ observation which occurred during Orbit 2 for all data taken during Orbit 1 and 2, and the \nustar\ observation taken during Orbit 3 for all data from Orbit 3. Then we fit all these 5 data groups simultaneously, without any stacking - Orbit 1 (no XMM), Orbit 1 (with XMM), Orbit 2 (no XMM), Orbit 3 (XMM), Orbit 3 (no XMM). Cross-calibration constants are used to account for calibration differences between the different instruments, as well as to account for variability in the hard energy band covered by the \nustar\ data.

\begin{figure*}
\begin{center}
\includegraphics[width=\textwidth]{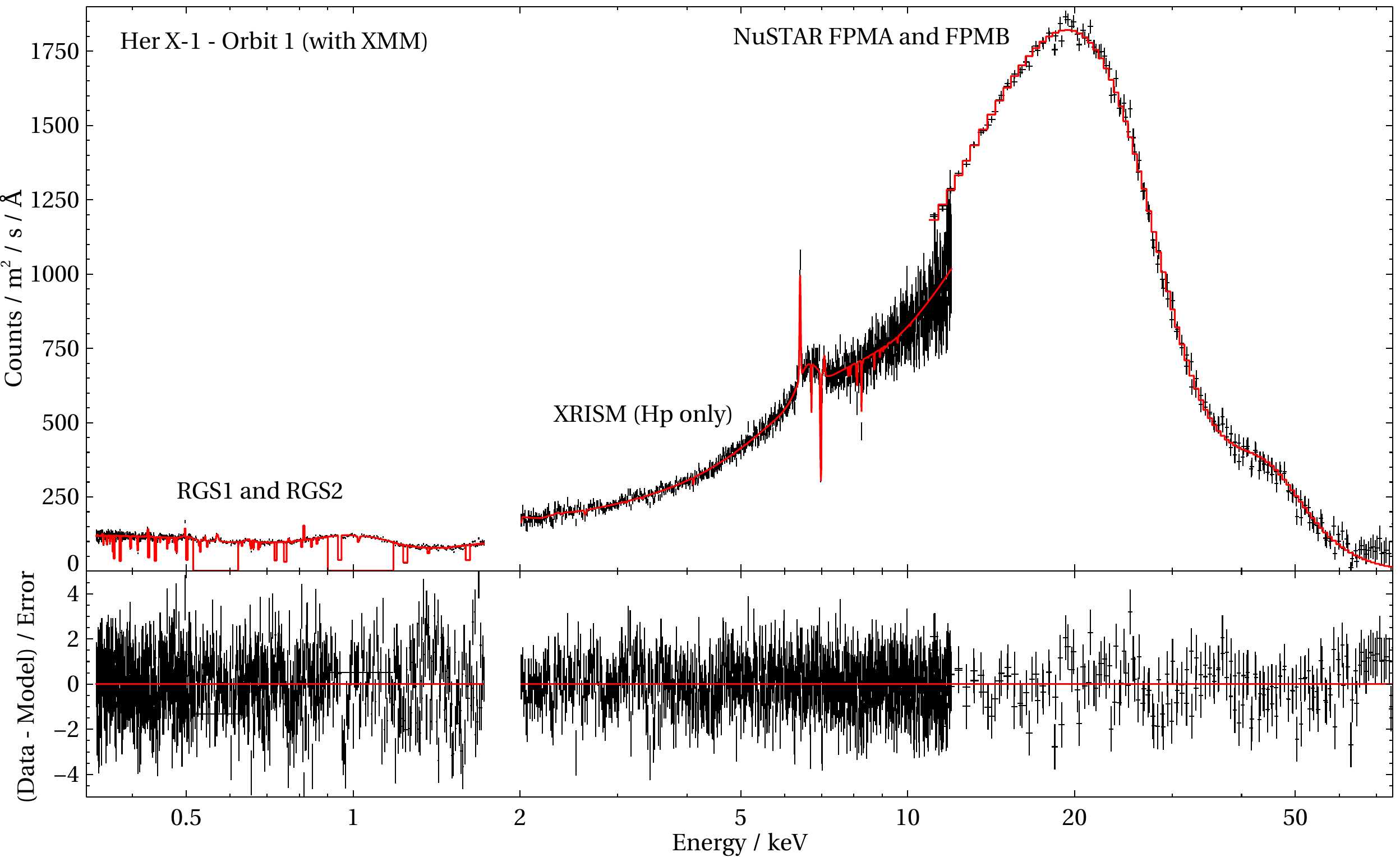}
\caption{The full-band spectral fit of Orbit 1 (with XMM) using the final \textsc{pion} model with the best-fitting disk wind elemental abundances, alongside with the residuals to this fit (lower panel). \xmm\ RGS 1, 2, \xrism\ (Hp events only), \nustar\ FPMA and FPMB data are shown. We note that most of the spikes across the RGS 1 and 2 model are not imprinted by the disk wind but are due to instrumental features (CCD chip gaps). \label{HerX1_SED}}
\end{center}
\end{figure*}

The spectral model used is similar to the one in section \ref{sec:slab}, but the phenomenological \textsc{slab} absorber model is replaced with a \textsc{pion} component. Moreover, we add several extra spectral components to describe the RGS data, and accurately describe the entire SED of Her X-1 from 0.35 keV to 75 keV. These spectral components are taken from \citet{Kosec+22}. The RGS spectrum is described with a combination of a regular blackbody (0.05 keV temperature, likely from direct disk emission), a blackbody modified with coherent Compton scattering (reprocessing of primary accretion column radiation with a temperature of $0.1-0.2$ keV), a broad Gaussian line near 1 keV (Fe L emission), two medium-width emission lines (O VIII and N VII emission) and 2 narrow emission lines (O VII and N VI emission). For those data groups with no XMM coverage, we link the RGS component parameters to the best-fitting values from the nearby data groups with XMM coverage. Considering the evolution of the \xrism\ lightcurve (Fig. \ref{xrism_lightcurve}), Orbit 1 (no XMM) has its RGS component parameters fixed to those from Orbit 1 (with XMM). Orbit 2 and Orbit 3 (no XMM) data groups have the RGS component parameters fixed to those from Orbit 3 (with XMM). One of these full-band spectral fits (Orbit 1 with XMM) is shown in Fig. \ref{HerX1_SED}.

All of the other spectral fit parameters are left free to vary (and are decoupled between the 5 data groups) except the elemental abundances in the \textsc{pion} component, which are all coupled among the 5 data groups. We note that all relevant elemental abundances across the \xmm\ and \xrism\ energy band are fitted (except the reference elements O or Fe) regardless of signal-to-noise, as leaving them fixed to 1 may bias the spectral fit. This very complex spectral model is then fitted in \textsc{spex} and we extract its statistical uncertainties. The best-fitting elemental abundances are given in Table \ref{el_abund} for both the O and Fe reference spectral fits. We note that the spectral fit quality of both approaches is nearly the same, and they differ by \delcstat\ of $\sim0.1$ for 44767 degrees of freedom. However, as expected, the best-fitting wind column densities \nh\ are different - setting Fe as the reference element results in 30\% lower \nh\ column densities in each data group, caused by increasing the strength of the lines of Fe (and other elements) by the same factor.

\begin{table}
\begin{center}
\caption{Best-fitting elemental abundances of the disk wind of Her X-1 from a simultaneous analysis of all \xrism, \xmm\ and \nustar\ data from the 2024 campaign. The first row shows the results of a spectral analysis assuming O as the reference element (the abundance of which is fixed to 1), while the second row shows the results of a spectral analysis assuming Fe as the reference element. \label{el_abund}}
\begin{tabular}{cccccccccccc}
\hline
\hline
Element& C & N & O & Ne &Mg & Si & S & Ar & Ca & Fe& Ni\\
\hline
\multirow{2}{*}{Abundances} & $0.00^{+0.08}$& $2.2_{-0.7}^{+0.8}$&1 (ref.)&$1.6_{-0.3}^{+0.4}$& $0.7_{-0.4}^{+0.5}$&$0.6_{-0.2}^{+0.3}$& $0.62_{-0.15}^{+0.17}$&$0.5_{-0.3}^{+0.3}$&$1.4_{-0.3}^{+0.4}$& $0.70_{-0.10}^{+0.08}$& $1.3_{-0.3}^{+0.3}$\\
&$0.0^{+0.1}$&$3.2^{+1.0}_{-1.0}$&$1.45_{-0.17}^{+0.19}$&$2.2_{-0.4}^{+0.5}$&$1.0_{-0.5}^{+0.7}$&$0.8_{-0.3}^{+0.4}$&$0.9_{-0.2}^{+0.2}$&$0.7_{-0.5}^{+0.5}$&$2.0_{-0.5}^{+0.5}$&1 (ref.)&$1.9_{-0.3}^{+0.3}$\\
\hline
\end{tabular}
\end{center}
\end{table}

The abundances of N and Ne are super-Solar compared with O, in agreement with previous measurements \citep{Kosec+23a}. The abundance of C was not previously measured in the disk wind but was found to be sub-Solar (C/O of $0.3-0.6$) in Her X-1 using Low state observations \citep{Jimenez+02}. Here we only obtained a very stringent upper limit of 0.08 on its value. The discrepancy with the study of \citet{Jimenez+02} could be introduced if there is a strong emission line component of C VI below the wind absorption, that may not necessarily be related to the disk wind. Due to low Doppler shifts of the wind, if there exists such a component, this would be very hard to separate, especially in RGS data. This same caveat in principle applies to all elemental measurements.

The abundances of Mg, Si, S and Ar are mildly sub-Solar, but have relatively large uncertainties. They are all consistent with the abundance of Fe at $1\sigma$ uncertainties. Our previous study using \chandra\ HETG \citep{Kosec+23c} indicated a super-Solar abundance of Mg, Si, and S, however our current results cannot be directly compared to that study, because in \citet{Kosec+23c} we fixed the abundances of both O and Fe to the best-fitting values from \citet{Kosec+23a}. On the other hand, the abundances of Ca and Ni are super-Solar, but again with significant uncertainties (abundances are consistent with Fe at $2\sigma$ level).

The biggest surprise is the abundance of Fe, which is tightly constrained to $0.70_{-0.10}^{+0.08}$. This result is in significant tension with previous \xmm\ results which indicated an abundance of $2.1 \pm 0.3$. We investigated this difference using the simultaneous \xmm\ EPIC-pn and \xrism\ data. To simplify the comparison as much as possible, we used the phenomenological \textsc{slab} model to fit the strength of the Fe XXV and XXVI lines. We found that EPIC-pn consistently overestimated the strength of these lines. If the Fe XXV/XXVI lines are overestimated in spectral modeling, the physical \textsc{pion} component will require a high \nh\ column density, or a high Fe abundance. This overestimation is likely due to the limited spectral resolution of EPIC-pn in the Fe K energy band, and a degeneracy between the optical depth of the wind absorption lines (at 6.7 and 7.0 keV) and the strength of the broadened Fe XXV emission line at the same location at 6.7 keV. Fixing the Fe abundance in the EPIC-pn data in a wind model using \textsc{pion} to the value from the current combined analysis (Fe $=0.7$) instead resulted in much higher column densities, underlying this disk wind-emission line degeneracy.

The spectral resolution of \xrism\ allowed us to finally resolve the individual features of the complex Fe K band of Her X-1, resulting in a much more confident measurement of the Fe line optical depths, and so a more confident measurement of the Fe/O elemental abundance ratio. It could be possible to further improve the quality of this elemental abundance study to include all of the previous high-quality \xmm\ (primarily RGS) data from the 2020 campaign as well as the \chandra\ HETG data. This would primarily reduce the uncertainties on the lighter elements within the RGS energy band (considering the wealth of available \xmm\ data) such as C, N, O, Ne and Mg. This is beyond the scope of the current paper as it would require more significant computational resources and is left for a future study.


\bibliography{References}{}
\bibliographystyle{aasjournalv7}



\end{document}